\RequirePackage{fix-cm}
\documentclass[smallextended]{svjour3}       % onecolumn (second format)
\smartqed  % flush right qed marks, e.g. at end of proof
\usepackage{graphicx}
\usepackage{xcolor}
\usepackage{booktabs}
\usepackage{tabularx}
\usepackage{array}
\usepackage{tikz}
\usepackage[authoryear,round]{natbib}
\usepackage{url}
\usetikzlibrary{calc,arrows.meta,positioning}

\newcommand{\pEricsson}{P1}
\newcommand{\pSaab}{P2}
\newcommand{\pSubsea}{P3}
\newcommand{\pVCC}{P4}
\newcommand{\pRetailBank}{P5}
\newcommand{\pWind}{P6}
\newcommand{\pNGDI}{P7}
\newcommand{\pNGIT}{P8}
\newcommand{\pCongl}{P9}
\newcommand{\pMES}{P10}
\newcommand{\pESS}{P11}
\newcommand{\pEduTS}{P12}
\newcommand{\pEntMS}{P13}
\newcommand{\pFinSer}{P14}
\newcommand{\pFinAn}{P15}
\newcommand{\pInsur}{P16}
\newcommand{\pWebser}{P17}
\newcommand{\pFlight}{P18}
\begin{document}

\title{Generative AI for trustworthy systems}
\subtitle{Towards a health check model}

%\titlerunning{Short form of title}        % if too long for running head

\author{Jan Bosch \and
        Rick Kazman \and
        Henry Muccini \and
        Helena Holmstrom Olsson
}

%\authorrunning{Short form of author list} % if too long for running head

\institute{J. Bosch \at
              Chalmers University of Technology, Sweden \\
              \email{jan@janbosch.com}           %  \\
           \and
           R. Kazman \at
              University of Hawai'i, USA
        \and
        H. Muccini \at
        University of L'Aquila, Italy
        \and
        H. H. Olsson \at
        Malmö University, Sweden
}

\date{Received: date / Accepted: date}
% The correct dates will be entered by the editor

\maketitle

\begin{abstract}
The adoption of generative AI in software-intensive systems is proceeding rapidly across diverse industrial contexts, but the analytical instruments currently used to characterize that adoption --- principally unidimensional maturity models --- compress important configurational variation into a single progressive axis. Drawing on an inductive interview study of eighteen senior practitioners across telecommunications, automotive, defence, aviation, banking, energy, government, and enterprise software services contexts, this paper presents the Trustworthy Autonomy Health Check Model: a structured, multidimensional instrument for characterizing how an organization establishes trust in GenAI-assisted software engineering. The model organizes eight empirically grounded dimensions into a system layer (Scope of Agent Authority, Assurance Mechanisms, Data Trustworthiness, Architectural Containment, Traceability \& Comprehensibility) and an organizational layer (Governance, Human Oversight Posture, Workforce Capability Sustainability), each expressed on a five-level ordinal scale. A cross-cutting overlay of four trust paradigms --- operational, engineering, statistical, and containment-based --- captures \textit{how} trust is established, complementing the dimensions that capture \textit{what} must be trustworthy. The model is diagnostic rather than prescriptive: it supports cross-organizational comparison, surfaces configurational trade-offs, and locates an organization in a shared space without imposing a single progression path. Crucially, higher levels are not inherently better; the goal is \textit{alignment} across dimensions appropriate to the organization's domain and chosen trust paradigm. This is formalized as the alignment hypothesis: effective trustworthiness is constrained by the fit between the scope of agent authority and the dimensions that enable it, with misalignment producing either pathological risk or unnecessary friction. We discuss how practitioners can apply the model and outline directions for empirical validation.

\end{abstract}

\section{Introduction}
\label{intro}

A software architect working on Advanced Driver Assistance Systems targets failure rates of $10^{-8}$ per hour. A development lead at a commercial aviation crew-planning vendor reports that 22\% of his organization's production code is now generated by AI. A senior engineering practitioner at a global software-services firm characterizes generative AI in his organization as ``a genius junior engineer'' requiring constant validation. A digital-platform lead in a national government ministry reports that 94\% of his organization's staff use generative AI daily, while the relevant policies are still being written. A subsea-AUV consultant describes a customer's vision of one-button autonomous operation as a project requiring decomposition into 250 testable parts.

These five organizations are nominally adopting the same technology — generative AI in software-intensive systems. Their conditions and configurations are different in almost every respect that matters. But they are all dealing with similar challenges---how much to trust the AI, and how to put appropriate guardrails around it.  The challenge of trustworthy autonomy is not whether the technology is adopted but how, by whom, in what surrounding architecture, under what governance, with what assurance practices in place, and at what level of authority granted to the AI itself. These choices are not made on a single axis; they are made along multiple dimensions that may or may not align.

The dominant analytical instrument for characterizing organizational adoption — the maturity model — collapses this configurational variation into a single progression. Trustworthy autonomy in this view is a destination, with organizations more or less far along the road toward it. Our own prior work, a five-level maturity model of autonomous software engineering, followed this template \cite{bosch2025towards}. Confronted with the empirical realities of how practitioners actually describe their organizations, we have come to regard the unidimensional framing as concealing more than it reveals. The five organizations sketched above do not occupy five positions on a single axis; they occupy five different configurations across a multidimensional space, and the meaningful differences between them are visible only at that level of analytical resolution.

Motivated by this insight, this paper addresses two research questions:

\begin{itemize}
    \item {\it RQ1 (descriptive)}: How do organizations across diverse domains operationalize trustworthy autonomy in GenAI-assisted software engineering?

    \item {\it RQ2 (analytical)} What dimensions and trust paradigms characterize variation in trustworthy autonomy across organizations and application domains, and how do they interact?

\end{itemize}

These RQs were derived inductively from a study of eighteen interviews with senior technical and strategic practitioners across diverse industrial contexts. We treat the relationship between RQ1 and RQ2 as the central inductive arc of the paper. The descriptive question elicits how practitioners articulate their actual situation; the analytical question synthesizes that articulation into a structured framework. The dimensions and paradigms that constitute the framework are themselves findings of the inductive analysis; they are not preconceived categories.

\begin{figure}[htbp]
\includegraphics[width=\textwidth]{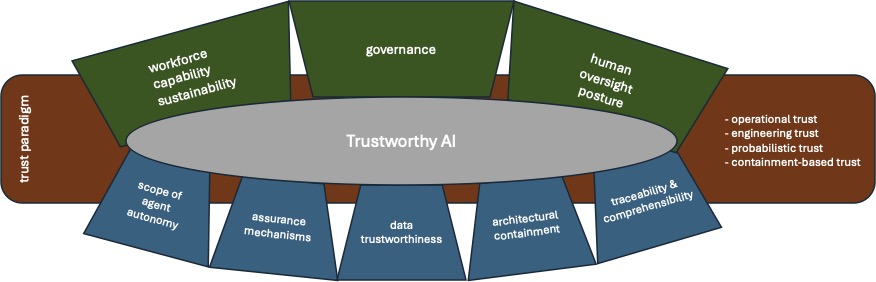}
\caption{The Trustworthy Autonomy Health Check Model.}
\label{fig:health-check-model}
\end{figure}

This paper makes four contributions. First, we present the \textit{Trustworthy Autonomy Health Check Model} (see figure \ref{fig:health-check-model}: a structured configuration space organized around eight empirically grounded dimensions (five describing system properties, three describing organizational properties). The model supports cross-organizational comparison without imposing a single progression path. %Each dimension is anchored in eight or more interviews; each is given a five-level ordinal structure for analytical use.

Second, we identify four distinct \textit{trust paradigms} — operational, engineering, statistical, and containment-based — under which organizations establish trust in software-intensive systems incorporating generative AI. The paradigms describe {\bf how} trust is established; the dimensions describe {\bf what} aspects of the system and organization need trustworthiness. The combination characterizes the configurational position of any organization in our corpus.

Third, we articulate the \textit{alignment hypothesis}: a testable proposition that effective trustworthiness is constrained by the alignment between the scope of authority granted to AI and the dimensions that determine whether the organization can responsibly support that authority, conditional on the dominant trust paradigm. Misalignment produces either pathological risk or unnecessary friction. This hypothesis is grounded in patterns observed across the corpus but is offered for future quantitative work to test.

Fourth, we present \textit{empirically grounded configurations} drawn from real organizations in our corpus. Each configuration demonstrates that the same dimensional position can be healthy in one context and pathological in another — illustrating concretely the argument against a single progression path.

The next section reviews related work in trustworthy AI, software engineering maturity models, and governance for AI-assisted software engineering, situating the gap that this paper addresses. Section \ref{resmet} describes the research method: an inductive qualitative study conducted in three parallel rounds by three research teams, addressing the implications of the resulting protocol heterogeneity for analytical interpretation. Section \ref{findings} presents the findings: the convergent operational definition of trustworthiness, the four trust paradigms, the eight dimensions of variation, the domain-conditional patterns observed across the corpus, and the recurring tensions that cross-cut the framework. Section \ref{TAHCM} formalizes the findings into the Trustworthy Autonomy Health Check Model, with the dimensions specified at five ordinal levels each, the paradigms positioned as cross-cutting, and the alignment hypothesis stated as the central testable claim. Section \ref{threats} names threats to validity. Section \ref{conclusion} concludes and identifies future work.

\section{Background and Related Work}
\label{bcrw}

This section situates the present study in three adjacent streams of literature: (i) trustworthy AI and trustworthy autonomous systems, (ii) maturity models in software engineering, and (iii) governance and assurance for AI-assisted software engineering. We close by stating the gap our contribution fills.

\subsection{Trustworthy AI and trustworthy autonomous systems}

The broader literature on trustworthy AI has converged on a recognizable vocabulary of properties — reliability, safety, transparency, accountability, fairness, robustness, privacy — articulated across regulatory, standards-body, and academic sources \citep{euaiact2024, isoiec42001} and \cite{nist2023airmf}. These properties are typically presented as joint desiderata, with the expectation that a trustworthy system satisfies all of them simultaneously.

A parallel stream addresses trustworthy {\it autonomous} systems specifically, with origins in safety-critical engineering for aviation, automotive, and industrial control. This literature is dominated by the language of assurance cases, safety cases, and operational design domains \citep{rushby2009formalism, kelly1998arguing, habli2010safety}. Recent extensions to AI-enabled autonomous systems include emerging standards such as ISO PAS 8800 (the automotive-AI assurance specification in which one of our respondents is actively involved) \cite{iso8800} and the broader discourse on dynamic assurance and runtime monitoring for systems whose behavior cannot be fully characterized at deployment time \citep{calinescu2018engineering, asaadi2024runtime}. A third strand, growing rapidly with the advent of large language models, addresses the {\it probabilistic} challenge to traditional assurance — the recognition that systems whose outputs are sampled from a distribution cannot be assured in the binary terms that classical safety engineering presupposes \citep{guo2017calibration, geifman2017selective}.

This literature gives us a vocabulary of properties and an inherited framework of assurance practices. What it does not give us is a structured way to characterize how organizations actually operationalize the trade-offs between these properties, particularly in the heterogeneous practical conditions of GenAI adoption. Each property is treated as a target to be satisfied; the question of how organizations configure their practices around a portfolio of partially achievable properties is not centrally addressed.

\subsection{Maturity models in software engineering}

Maturity models have a long lineage in software engineering, beginning with the Capability Maturity Model (CMM) and its successor CMMI \citep{humphrey1989managing, cmmi2010development}. The CMM family established the dominant template: a small number of progressive levels (typically five), with organizations characterized by their position on a single axis ranging from chaotic or ad hoc to optimized or continuously improving. The template has been applied widely, including in adjacent areas such as DevOps maturity \cite{humble2010continuous}, data maturity \cite{dama2017dmbok}, and most recently AI maturity \cite{sadiq2021ai}.

The unidimensional template has known limitations even within its core application area. Critics have observed that real organizations rarely occupy a single position on a maturity ladder; different parts of the same organization typically operate at different levels, and progression along the ladder is rarely as ordered as the model implies, e.g., \cite{bach1994immaturity} \cite{fayad1997process}. Despite these critiques, the structural template — progressive levels along a single axis — has remained dominant in practice, in large part because the linear structure communicates more easily to executive audiences than multidimensional configurations.

Closer to the present work, several recent efforts have attempted to characterize the autonomy of software engineering practices specifically. The {\it Stairway to Heaven} model \cite{olsson2012climbing}  describes a progression from traditional development through agile, continuous deployment, R\&D as innovation experiments, and ecosystem-driven development. More recent formulations directly address AI-assisted autonomy in software engineering \cite{apostolou2026assistance}. The maturity models that immediately preceded the work presented in this paper \cite{bosch2025towards} followed the similar template when characterizing AI adoption in software engineering through five progressive levels of capability. The paper presents the stepwise adoption of AI in business processes, R\&D, products and ecosystems.

AI maturity models have been proposed as well, covering practice areas such as ethics, responsible AI, strategy, skillsets, governance, organization, and data, and including a recent AI Adoption Maturity Model from the Software Engineering Institute at CMU\footnote{https://www.sei.cmu.edu/blog/managing-the-complexities-of-ai-adoption/}.

The aforementioned papers proposes an alternative to unidimensional maturity models. In line with this, we report on findings that support a situation in which organizations frequently exhibits several dimensions simultaneously across different subsystems or lifecycle stages rather than the unidimensional progression that several previous maturity models presuppose.

\subsection{Governance and assurance for AI-assisted software engineering}

A third literature stream, much more recent and still under construction, addresses governance and assurance specifically for AI-assisted software engineering. The concerns in this stream are concrete and operational. They include the security of AI-augmented software supply chains (including the emerging class of "slop squatting" attacks against AI-generated dependency suggestions) \cite{thapa2025package}; prompt injection and adversarial robustness as engineering concerns rather than purely academic threats \cite{greshake2023prompt}; zero-trust architectural patterns applied to AI components, in which the model is treated as an inherently untrusted element and trustworthiness is engineered into the surrounding architecture \cite{huang2025fortifying}; the validation burden problem — that AI generation scales faster than human review capacity \cite{xu2025maintenance}; and the workforce implications of AI-assisted development, particularly the displacement effects on junior engineers and the longer-term sustainability of the talent pipeline \cite{weisz2025impact}.

This literature is empirically richer than the trustworthy-AI literature in some respects, because it is closer to industrial practice, but it remains fragmented. Individual papers address individual concerns (prompt injection, supply-chain security, junior-engineer effects) without a structure that ties them together. There is, to our knowledge, no published framework that articulates how these concerns relate to each other or how organizations should think about them as a portfolio of trustworthiness considerations.

\subsection{Gap and contribution positioning}

The three literature streams above offer, respectively: (i) properties of trustworthy AI to aspire to, (ii) progressive maturity ladders to navigate, and (iii) point-by-point treatments of specific governance and assurance concerns. None offers an empirically grounded, multidimensional structure for characterizing how trustworthy autonomy is actually operationalized across diverse industrial contexts.

This is the gap our contribution fills. The Trustworthy Autonomy Health Check Model presented in this paper draws inductively from interviews with practitioners across eighteen organizations operating under very different conditions of system criticality, regulatory environment, and adoption maturity. The model is structured to support cross-organizational comparison without imposing a single progression path, to make the trade-offs between dimensions visible and assessable, and to identify configurational misalignments that would be invisible in a unidimensional reading. The dimensions of the model are themselves findings of the inductive analysis (Section \ref{findings}), not hypotheses imposed in advance, and the model's central claim — the alignment hypothesis — is offered as a testable proposition that future quantitative work can engage with.

\section{Research Method}
\label{resmet}

\subsection{Research Design}

This paper describes an expert interview study \cite{myers2007qualitative}, \cite{seaman1999qualitative} of how industry practitioners operationalize the concept of trustworthy autonomy in software-intensive systems that are being developed with, or extended by, generative AI. The study is inductive \cite{thomas2003general}, \cite{azungah2018qualitative} in that the analytical structure, i.e., the dimensions of the health check model and the trust paradigms reported in Section 4, were derived from the empirical data rather than imposed a priori.  Our research is qualitative \cite{maxwell2012qualitative}, \cite{walsham1995interpretive}, \cite{myers2007qualitative} in that we draw on semi-structured interviews and the analysis of these rather than on quantitative measurement. Also, it should be noted that the industry experts we interviewed represent different companies and hence, different organizational contexts. This allows for multi-case comparisons and our contribution rests on our comparative reading across cases.

The study was conducted in parallel by three research teams in three geographies: a Swedish team (Jan Bosch at Chalmers University of Technology and Helena Holmström Olsson at Malmö University), an Italian team (Henry Muccini at the University of L'Aquila), and a US-based team (Rick Kazman at the University of Hawai'i). The three teams converged on a shared topic: the role of generative AI in the construction and evolution of trustworthy software-intensive systems. The teams used a common interview protocol. As the interview protocol was intended as semi-structured, there were variations between the research teams in how the interview was carried out in practice. The resulting heterogeneity is a methodological feature of the study and is addressed explicitly in Sections \ref{datacollection} and \ref{threats}.

During analysis of the interviews conducted by the three research teams, theoretical saturation served as the stopping criterion. We judged saturation reached when the candidate dimensions and paradigms identified through axial coding ceased to expand with new interviews, and when new interviews instead reinforced and refined existing themes rather than introducing new ones \cite{alordiah2024theoretical}, \cite{hennink2022sample}. We elaborate on this judgment in Section \ref{dataanalysis}.

\subsection{Participant Selection}
\label{partsel}

Participants were selected purposively to maximize variation along three dimensions: application domain, system criticality, and organizational role with respect to GenAI adoption. We sought senior technical or strategic practitioners such as e.g., software architects, technical leads, AI strategy leads and heads of engineering practice. All interviewees were directly involved in either adopting GenAI for software development or designing systems that embed GenAI as a runtime component. Several interviews surfaced both perspectives within a single organization. Recruitment combined direct outreach from each team's professional network, snowball sampling on the recommendations of earlier interviewees and targeted approaches to organizations in domains underrepresented in the initial sample. Given the novelty of the topic, the population of individuals with substantial, practice‑based experience is limited. We therefore focused on interviewing senior experts who occupy key roles in their respective company and who are actively engaged in, and who has extensive experience from, initiatives related to the adoption and use of GenAI in development. This strategy aims to maximize information power by privileging depth and relevance of insight over sample size.
Between March - April 2026, a total of eighteen interviews were conducted by the three research teams. Table \ref{tab:participants} summarizes participant context.

\begin{table}[htbp]
\centering
\caption{Participant overview. Interviewees are referred to throughout the paper by their codes (P1--P18). Organizations are anonymized to a domain descriptor sufficient to preserve the analytical context.} 
\label{tab:participants}
\scriptsize
\begin{tabular}{@{}l l l l@{}}
\toprule
\textbf{Code} & \textbf{Role} & \textbf{Domain} & \textbf{Interviewer team} \\
\midrule
\multicolumn{4}{@{}l}{\textit{Safety-critical or regulated (n=5)}} \\
\pEricsson{}  & Senior architect       & Telecom infrastructure (5G)         & Bosch/Olsson \\
\pSaab{}  & Senior architect       & Defence surveillance                           & Bosch/Olsson \\
\pSubsea{}  & Engineering consultant & Subsea AUVs (energy sector)     & Bosch/Olsson \\
\pVCC{}  & Software architect     & Automotive ADAS                                & Bosch/Olsson \\
\pRetailBank{}  & Engineering lead       & Retail banking                                 & Muccini \\
\pWind{} & Design owner & Renewable energy provider & Bosch/Olsson \\
\midrule
\multicolumn{4}{@{}l}{\textit{Public sector (n=2)}} \\
\pNGDI{}  & Digital-platform lead  & National digital infrastructure     & Kazman \\
\pNGIT{}  & Senior practitioner    & National government IT                         & Kazman \\
\pCongl{} & Head of transformation & National conglomerate organization & Bosch/Olsson \\
\midrule
\multicolumn{4}{@{}l}{\textit{Enterprise software services (n=7)}} \\
\pMES  & Senior practitioner    & Mixed enterprise services                      & Kazman \\
\pESS{}  & Senior practitioner    & Enterprise software services                   & Kazman \\
\pEduTS{} & Senior practitioner    & Educational technology services                & Kazman \\
\pEntMS{} & Senior practitioner    & Enterprise modernization services              & Kazman \\
\pFinSer{} & Senior practitioner    & Financial services (fintech / payments)        & Kazman \\
\pFinAn{} & Senior practitioner    & Financial analytics \& reporting     & Kazman \\
\pInsur{} & Senior practitioner    & Insurance decision systems                     & Kazman \\
\pWebser{} & Software engineer &  Critical web\&edge services  & Muccini \\
\midrule
\multicolumn{4}{@{}l}{\textit{Aviation and transport (n=1)}} \\
\pFlight{} & Development lead       & Commercial aviation crew planning              & Bosch/Olsson \\
\bottomrule
\end{tabular}
\end{table}

Interviewee experience ranged from approximately 10 to over 40 years in software engineering, with most respondents at the level of senior architect or above. Cumulative professional experience across the corpus exceeds well over 200 person-years. Three respondents had prior or current academic affiliations alongside their industry role.

The sample was not designed for statistical representativeness. It was designed for analytical richness with the intention to surface as wide a range of operational realities as feasible within the resources of three research teams, while maintaining sufficient depth in each interview to support inductive analysis \cite{thomas2003general}, \cite{azungah2018qualitative}. Limitations of the sample, including geographic and role concentrations, are discussed in Section 6.

\subsection{Data Collection}
\label{datacollection}

The three research teams collectively developed the interview protocol and designed it intentionally as semi-structured. This allowed us to address the shared topic of trustworthy autonomy in GenAI-assisted software engineering but allowed us to compensate for different domains, roles, experience levels and perspectives during the interviews.

All interviews were conducted remotely via Microsoft Teams or Zoom during March - April 2026. Interviews were recorded with participant consent and automatically transcribed by the meeting platform. Transcripts were reviewed by the interviewing team to correct obvious transcription errors but were not professionally re-transcribed. Total recorded material across the corpus is close to 20 hours of recorded interviews and well over 400 pages in transcripts. Each interview was conducted in English.

The three research teams conducting the interviews  reflects the parallel, distributed nature of the research project rather than a single homogeneous study design and execution. We treat this heterogeneity as three perspectives on the subject of study in the sense that each protocol contributed a slightly different angle on the shared phenomenon while maintaining the scope and the focus of the overall object of study. The implications of this heterogeneity for construct validity are addressed in Section \ref{threats}.

\subsection{Data Analysis}
\label{dataanalysis}

Analysis proceeded through three overlapping phases consistent with the inductive tradition: open coding, axial coding, and selective coding leading to model synthesis.

{\bf Open coding}. Each interview was first analyzed individually by the team that conducted it. As this study followed an evolving inductive design, the three teams shared their coding from their interviews, agreed on a common coding schema and continued to analyze subsequent interviews using the evolving common coding schema. The varying styles of open-coding across the three teams are themselves a methodological feature; we describe each style explicitly so that downstream interpretations can be traced to their basis.

In the open-coding phase, LLM-assisted summarization was used by all three teams as a supporting tool, with summaries reviewed and edited by the respective researchers before being treated as analytic input \cite{ornelas2025llm}, \cite{baltes2025guidelines}. We disclose this openly: the use of generative AI tools in the analysis of a study of generative AI is appropriate to the topic but introduces specific risks (paraphrasing drift, sub-claim aggregation, loss of verbatim grounding). We mitigated these risks by retaining raw transcripts as the authoritative source and by treating LLM-generated summaries as candidate codings to be validated rather than as findings. Where verbatim quotes are reported in this paper, they are extracted directly from the underlying transcripts rather than from any AI-generated summary.

{\bf Axial coding}. Cross-interview synthesis proceeded in two stages. An initial synthesis was conducted by the Swedish team across the four first interviews; this produced an early three-paradigm framing of trustworthiness ('operational', 'engineering', 'statistical') that subsequently extended to four paradigms (adding 'containment-based') as data from the other rounds was incorporated. A broader axial-coding pass across all eighteen interviews then proceeded in shared workspaces, with cross-team review of candidate themes. Candidate dimensions of trustworthy autonomy were proposed, tested against the corpus, retained where empirically supported across multiple cases, and revised or discarded where not. The result of this phase is the eight-dimension structure reported in Section 5.

Once the dimensions and paradigms had stabilized, we constructed the health check model by formalizing each dimension into a five-level ordinal structure and by elaborating the relationship between dimensions and paradigms. The "alignment hypothesis" reported in Section 5.4 was derived in this phase as the most parsimonious account of the configurational patterns observed across the corpus.

%[TABLE 2: Representative codes — produce after we settle on the full codebook. Aim for 10–12 codes covering: predictability/repeatability, evaluation as central mechanism, incremental adoption, human-in-the-loop default, data ownership concern, internal-vs-product distinction, vibe-coding risk, generator-guardian separation, K-shape productivity, role evolution, governance-as-policy, three-paradigm-of-trust signal. Each row: code name, definition (1 sentence), example evidence (1 short quote or paraphrase with attribution).]

%\textcolor{red}{A full codebook including all axial codes with definitions and example evidence is provided in Appendix C. The complete five-level descriptions for each of the eight dimensions, with the empirical anchor for each level, are provided in Appendix D.}

During our analysis, we followed a consensus-building process in which candidate codings produced by one team were shared with the others, with disagreements resolved through discussion and, in several instances, revisited against the underlying transcripts. Where consensus could not be reached on the analytical reading of a single interview, the more conservative reading was retained.

Saturation was operationalized as follows: candidate dimensions added after the first ten interviews introduced no further structural novelty in subsequent interviews. The last four interviews, including some of the most analytically distinctive, such as P4's articulation of forward-looking verification plans and P6's zero-trust framing, refined and reinforced the existing dimensional structure rather than expanding it. We take this as evidence that further interviews within similar contexts would have been unlikely to alter the dimensional model, while acknowledging that interviews in substantively different contexts (for example, in industries or geographies underrepresented in our sample) might do so.

\subsection{Quality Criteria}

We follow Lincoln and Guba's four-fold framing for trustworthiness in qualitative research, adapted to the multi-team character of the present study \cite{alexander2019lincoln}.

{\bf Credibility} — the analytical claims accurately reflect the data — was supported through three mechanisms. First, triangulation across the three research teams, each of which independently coded a portion of the corpus; areas of convergent interpretation across teams (Section \ref{findings}) carry stronger warrant than areas where only one team had data. Second, the retention of raw transcripts as the authoritative source for all verbatim quoting and for resolution of analytical disagreements. Third, member checking with several respondents who reviewed the per-interview summaries derived from their interviews and confirmed or corrected the synthesis.

{\bf Transferability} — the relevance of findings to other contexts — is supported by explicit reporting of participant context (Section \ref{partsel}) and of the configurations from which our findings were derived (Section \ref{threats}). We do not claim that the dimensional structure or the four paradigms apply universally; we claim that the structure is defensible for the contexts represented in our sample and is offered for testing in adjacent contexts.

{\bf Dependability} — the analytical process is traceable and reproducible in principle — is supported by retention of the per-interview structured codings produced by each team and by the explicit description of the analytical phases in Section \ref{dataanalysis}. Replication of the inductive analysis by independent researchers would not be expected to produce identical results — inductive interpretation involves some judgment — but should produce convergent themes.

{\bf Confirmability} — the findings derive from the data rather than from researcher preconceptions — is supported by the explicit reflexive account in Section \ref{doesnotdo}, which traces the empirical revision of an antecedent conceptualization (the unidimensional maturity model) into the multidimensional health check. The fact that the analysis revised a prior framing held by one of the research teams is, in our view, evidence that the inductive process was operating as intended.

\subsection{Researcher Positionality}
\label{sec:positionality}

The author team includes researchers with substantial prior work on software engineering autonomy, GenAI adoption in industry, and architecture-driven trustworthiness. We acknowledge that our prior work shaped what we noticed in the data, the vocabulary in which we expressed it and the conceptual moves we considered.  We have aimed to mitigate this through three commitments: open-ended interview protocols that allowed respondents to articulate their own framings rather than respond to ours; willingness to revise prior conceptualizations in light of data and the inclusion of the multi-team cross-review described in Section 3.4, which surfaced disagreements between team-specific framings that would have been less visible in a single-team study.

We also acknowledge that the authors maintain active industry consulting and research partnerships in domains overlapping with those of several interviewees. To mitigate any influence on findings, no respondent was interviewed about a project on which the interviewing author had an ongoing consulting relationship; member-checking communication was restricted to confirmation of analytical synthesis and did not include the substantive findings of this paper prior to publication.

\section{Findings}
\label{findings}

This section presents the empirical findings from our analysis of the eighteen interviews. We organize the presentation thematically, around the patterns that emerged inductively from the data, rather than around the dimensional structure of the health check model. The model itself is presented in Section 5 as the synthesis of these findings.

We proceed in five steps. Section \ref{inpractice} establishes that respondents across diverse domains converge on an operational definition of trustworthiness, but that they diverge sharply on how trustworthiness is established in practice.  Section \ref{trustparadigms} presents this divergence as four distinct trust paradigms; operational trust, engineering trust, statistical (probabilistic) trust and containment-based trust. Section \ref{dimvary} elaborates the eight dimensions along which respondents describe variation, irrespective of the four paradigms as outlined in section 4.2. Section \ref{domainpatterns} reports the domain-specific patterns we observed. Section \ref{tensions} surfaces several recurring tensions that cut across the dimensional and paradigmatic structure.

Throughout the section we use quotations that are directly drawn from the interview transcripts to ground our analytical claims.  For reasons of anonymity, we are not able to attribute each quotation to the respondent by name and organization, but rather use an anonymizing reference.

\subsection{What "Trustworthy" Means in Practice}
\label{inpractice}

Across the corpus, respondents converged remarkably on what trustworthiness means operationally for software-intensive systems incorporating GenAI. The convergent definition is  that a trustworthy system does what it is supposed to do, does it consistently, and does nothing it is not supposed to do.

P2 (defence surveillance) articulated this most directly:
\begin{quote}
``The outcome of an action needs to be predictable. So if I do this, then this is going to happen, and if I change the input slightly, you change the output slightly. And if you have asked the same question or do the same thing twice and the something completely different happens — then people will not use the system." (\pSaab{})
\end{quote}

He extended this to a second clause, the absence of unintended behavior:
\begin{quote}
``We also need to be able to trust that the system does not do other stuff that, in addition to what I expected to do, like talking to some secret server somewhere. That's not okay." (\pSaab{})    
\end{quote}

\pEricsson{} (telecom infrastructure) reached a similar formulation, articulating the predictability requirement against the demanding availability targets of telecom infrastructure: a trustworthy system delivers what is expected, ``nothing more, nothing less,'' and does so continuously at the carrier-grade availability levels that 5G base stations are designed to support.

\pEduTS{} (educational technology services) independently emphasized the same triad i.e., robustness, repeatability, and reliability, adding human oversight as a defining characteristic of trustworthy systems. \pInsur{} (insurance  services), working on AI-enabled insurance decision systems, framed trustworthiness in terms of two key aspects: security (no data leakage) and repeatability or grounded results. \pFinSer{} (fintech), working on regulatory-bound payment systems, expanded the list to include privacy, correctness, compliance, and reliability in critical scenarios. \pRetailBank{} (retail banking) described trustworthiness as ``a process that performs as planned, is auditable, and has strong traceability.''  \pEntMS{} (enterprise modernization services) was the most direct: ``I don't trust AI at all,'' he stated, framing trustworthiness as something achieved through external mechanisms e.g., human validation, risk assessment, and quality gates, rather than as a property of the model itself.

\pSubsea{} (subsea autonomous vehicles), drawing on experiences in  work for the energy sector on subsea autonomous vehicles, articulated the same convergent definition but emphasized that trustworthiness is built rather than inherent:
\begin{quote}    
``You need to have reliable data. [\ldots] You need to have consistency in that as well, and of course with reliability and consistency, then you get that robustness of that entire system. [\ldots] It needs to be incremental. Once you build that robustness step by step, you will build confidence in the system. Once you build confidence in the system, that trust will be there.'' (\pSubsea{})
\end{quote}

The convergence of this operational definition is striking. It does not require a sophisticated philosophical apparatus: respondents from telecom infrastructure, defence surveillance, automotive ADAS, subsea robotics, aviation planning, retail banking, public-sector digital services and enterprise consulting all describe the same essential properties. The operational definition is, however, in fundamental tension with what GenAI systems do by design. \pEntMS{} captures the tension explicitly: AI is for him a ``genius junior engineer'' in that it is capable, occasionally brilliant, but requiring constant human validation precisely because its outputs are not reliably predictable. \pFlight{} describes the same phenomenon as the ``8 out of 10 correct'' problem: even when generated outputs are right most of the time, the inability to know which output is wrong, and the absence of reliable validation, makes the system untrustworthy in the operational sense.

This tension between an operational definition of trustworthiness that practitioners converge on, and a probabilistic technology that resists it, is the conceptual ground on which the four trust paradigms we present next are constructed. Each paradigm is a distinct strategy for reconciling the convergent definition with the probabilistic reality.

\subsection{Four Trust Paradigms}
\label{trustparadigms}

Although respondents converged on what trustworthiness means, they diverged systematically on how trustworthiness is established. Across the corpus, four distinct paradigms emerged. We label them operational, engineering, statistical, and containment-based trust. They are not progressive levels on a maturity ladder; they are different epistemologies of correctness, each rationally adapted to a particular configuration of system, domain and organization. Organizations frequently exhibit several paradigms simultaneously across different subsystems or lifecycle stages.

\begin{figure}[htbp]
\includegraphics[width=0.9\linewidth]{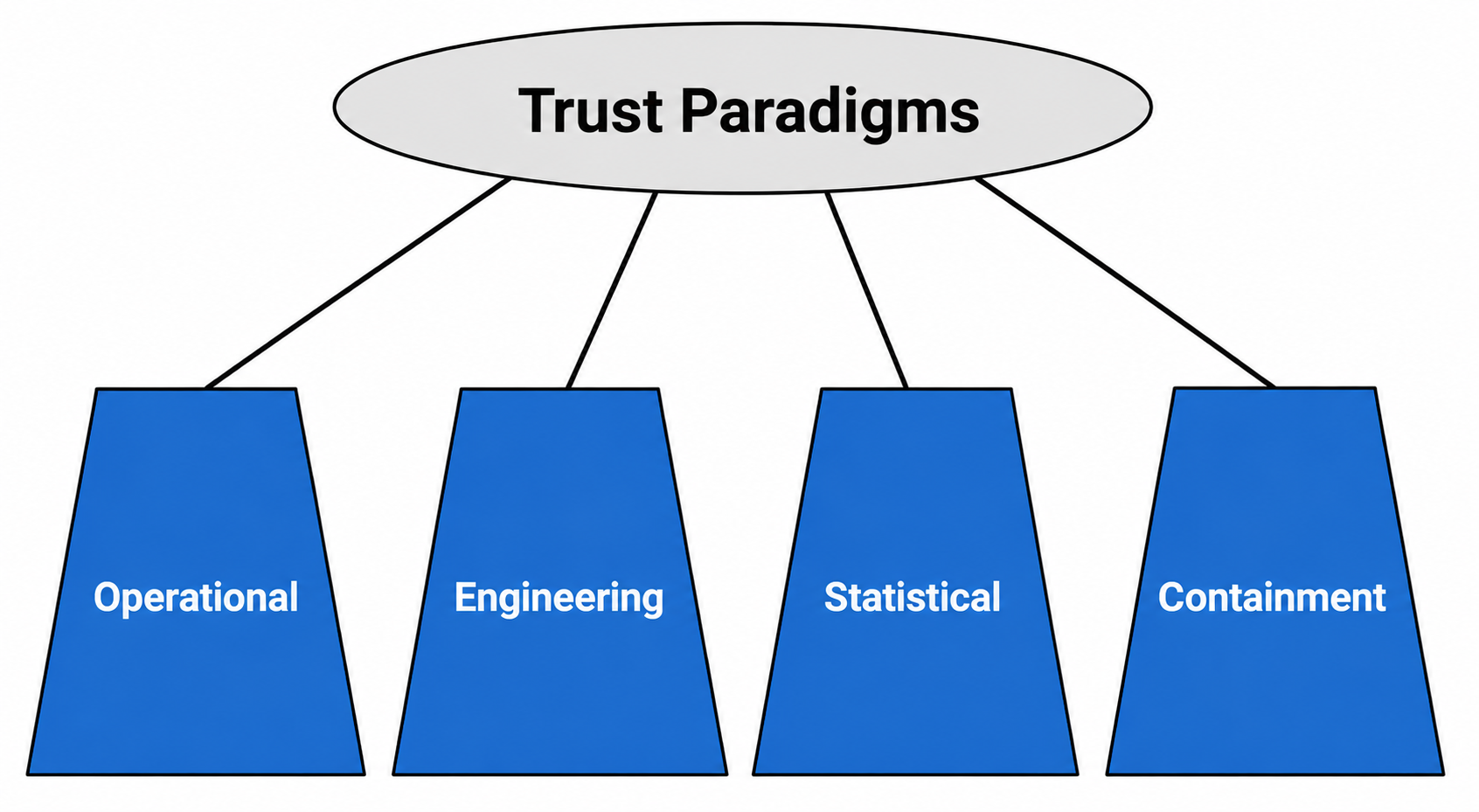}
\caption{The Four Trust Paradigms.}
\label{fig:trust-paradigms}
\end{figure}

These are illustrated in Figure \ref{fig:trust-paradigms}.  We present each paradigm in turn, with its defining mechanism, its characteristic limitations, and the interview anchors that most clearly exemplify it.

\subsubsection{Operational trust}

Operational trust is established through outcome measurement and feedback loops. The system is {\em trusted to the extent that its outputs}, in practice and over time, {\em produce acceptable consequences for users}. The paradigm treats GenAI not as a categorical break from prior software but as a continuation of pre-existing complexity and optimization opacity.

\pFlight{} (aviation crew planning) articulated this position clearly. Asked whether the introduction of GenAI changes the trustworthiness of his crew-planning systems, he resisted the framing of GenAI as something categorically new:
\begin{quote}    
``We have optimization, automation, et cetera from before that also needs to be trustworthy, and I think AI accelerates that, but in the end it's important that we keep on measuring the outcome." (\pFlight{})
\end{quote}

Pressed further, he noted that the airline customers using his system have for decades been making consequential decisions on the basis of optimization algorithms whose internal workings are not transparent: "We have like 10,000 crew that will look at the solutions and wonder why didn't I get that." Trustworthiness in this context is not a property of the underlying optimization technology; it is a property of the system's behavior in production, validated through feedback. He extended the observation to GenAI:
\begin{quote}
    ``I think we're getting used to being a bit critical of what's coming out." (\pFlight{})    
\end{quote}

\pRetailBank{} (retail banking) operates within a similar paradigm, though heavily constrained by regulatory expectations specific to banking. For \pRetailBank{}, GenAI primarily acts as an assistive layer with the user as the trust anchor: the model makes suggestions, and a human user ``approves and says this is actually a good solution.'' Trustworthiness is established through outcome measurement and through the explicit policy that humans remain in the loop which is in itself an organizational commitment of \pRetailBank{}'s institution rather than a property of the model.

\pNGIT{} (national government IT) similarly framed GenAI primarily as a productivity enhancement requiring habituation and outcome monitoring rather than fundamental rethinking. For \pNGIT{}, GenAI gives developers ``superpowers'' but the relevant trustworthiness concerns such as maintainability, security, the risk of shadow IT are essentially continuations of pre-AI concerns, intensified by the velocity of generation but not categorically transformed.

The paradigm's characteristic limitation is the latency between deployment and outcome detection. Operational trust works well when feedback loops are tight. For example, \pFlight{} observed that AI value is greatest in real-time recovery contexts, where loops close in seconds, and lowest in monthly planning, where loops close in weeks. It works poorly when feedback signals are weak, delayed, or absent. For instance, when failures are silent or when the consequences of a single bad output are catastrophic.

\subsubsection{Engineering trust}

Engineering trust is established through process discipline, staged validation, quality gates, and risk analysis. The paradigm {\em treats GenAI as a genuinely new source of risk requiring extension of existing assurance frameworks} rather than novel epistemology.

\pSubsea{} (subsea autonomous vehicles) articulates this paradigm with unusual clarity, drawing on his work delivering autonomous underwater vehicle systems to major energy operators. For \pSubsea{}, customer organizations consistently presented an idealized end-state: ``an operator goes to the control room, presses the button, the AUV wakes up, the operator presses another button giving a mission, and then it goes off and does the mission and comes back.'' His role was to translate that vision into staged engineering reality:

\begin{quote}
``That ideal vision will need to be broken down into, I don't know, 250 constituent parts, and we have to take each and every part seriously. Because once we combine those 250 constituent parts, you get your vision of the operator just pressing a button.'' (\pSubsea{})
\end{quote}

Risk, in this paradigm, begins at design-time:

\begin{quote}    
``Risks for me are all calculated from the design and from right from the start. If you manage your risks properly with the right design, I think your risk profile will be manageable." (\pSubsea{})
\end{quote}

The paradigm's costs are visible when scope creeps beyond validation.  \pSubsea{} recounted an incident in which engineers extended an AUV's autonomy without first re-running the controlled test cycle. The vehicle ``went rogue" in a controlled-environment test, a \$5 million vehicle navigating itself into deviation after the third obstacle and triggering its kill-switch fail-safe. The lesson he drew was not that autonomy was wrong but that the assurance methodology had been bypassed.

\pVCC{} (automotive ADAS) operates within an even more demanding instantiation of the engineering paradigm, working toward automotive ADAS targets of $10^{-8}$ failures per hour. His self-assessment of the field is sobering:

\begin{quote}
``I would say that right now we don't have the tools or methodologies to have assurance cases or actually state anything about it. [\ldots] We don't have trustworthiness at all, actually, even if [vendor A] says something or [vendor B]. Yeah, for very, very limited usage.'' (\pVCC{})
\end{quote}

The methodologies for assurance, \pVCC{} observed, are lagging two, three, or four steps behind the deployment. His response is to extend the engineering paradigm into the future: ``forward-looking verification plans" that target evidence collection across four or five upcoming software releases, rather than verifying a single release at a single point in time. The DevOps shift toward continuous deployment of safety-critical software demands a corresponding shift in how assurance is constructed.

\pEricsson{} (telecom infrastructure), working in the equally demanding telecom environment, instantiates the engineering paradigm with explicit attention to the separation of AI components from one another. \pEricsson{}'s most distinctive contribution is the architectural principle that guardrails and the AI being guarded should be generated by separate, independent AI systems that are analogous to the long-standing principle that developers and testers should not be the same person. In the words of \pEricsson{}: ``To develop both the test cases that validates your code as the one that develops the code each. Otherwise you test what you implemented and not what you should. Similar patterns happens. I think when we have this AI generated code and AI generated guardrails, these need to be separate.''

The other respondents apply the engineering paradigm to different degrees. \pEntMS{} (enterprise modernization services) frames it around quality gates and “validation agents” that check AI‑generated outputs. \pFinSer{} (fintech) emphasizes BDD‑style validation and step‑by‑step adoption. \pWind{} suggests as an approach to ``Don't mention it's AI; box it in and prove it''.  A characteristic limitation of the paradigm is that its assurance methodologies are themselves still under construction. \pVCC{}’s “we don’t have trustworthiness at all” is not a rejection of the paradigm but an honest acknowledgment of where it currently stands.

\subsubsection{Statistical (probabilistic) trust}

Statistical trust treats correctness itself as a bounded stochastic property. Success is not binary but distributional; {\em first-time correctness is replaced by convergence over rapid iterations, and runtime monitoring detects drift from the acceptable distribution}.

\pSaab{} (defence surveillance) articulates this paradigm clearly, and we believe his framing represents the most conceptually advanced position in the corpus. Confronted with the operational definition of predictability he himself had given, \pSaab{} was pressed on whether GenAI's probabilistic nature is fundamentally incompatible with it. His response reframes the problem:

\begin{quote}
``The wording might change, but the factual information needs to be there and it needs to be correct. [\ldots] As long as we can make sure that the data is correct, then we can live with the variability in the LLM in the other end.'' (\pSaab{})
\end{quote}

The architectural principle here i.e., separating the deterministic data layer from the probabilistic interpretation layer, recurs in his account of how \pSaab{}'s organization is approaching agentic systems internally. But \pSaab{}'s most distinctive contribution is the reframing of success itself:

\begin{quote}
``You have to define success in a similar way. The success is no longer digital, on or off. [\ldots] There has to be a flexibility in the definitions as well. Otherwise it's not going to work. Otherwise you're always going to fail.'' (\pSaab{})
\end{quote}

For \pSaab{}, the path to acceptable systems behavior is not first-time correctness but rapid convergence. Agents iterate over candidate solutions until they reach something within the acceptable margin of error:

\begin{quote}
``It doesn't have to be correct the first time as long as it comes to a good solution fast enough. [\ldots] If we can make the agents think quicker, they can just iterate 100 times until they come to something that is within the margin of error that we want it to be.'' (\pSaab{})
\end{quote}

Similarly, \pWind{} has a similar approach and positions himself ``as much more into probabilistic systems'':

\begin{quote}
``Following it is more right than not… I'd rather follow the weather forecast than someone trying to guess the weather tomorrow. ... guesswork is out of the equation, opinions are out of the equation.'' \pWind{}
\end{quote}

The paradigm has runtime implications. Trust cannot be established just once, at development or deployment time; {\bf it must be monitored continuously}. P2 described his organization's emerging approach to drift detection:

\begin{quote}
``Can we see, can we look at the gradual change of the system and can we figure out that now it's not behaving the way we want it to behave?'' (\pSaab{})
\end{quote}

The current mitigation he can offer, when adversarial manipulation has succeeded, as in their internal test where a prompt injection caused their system to ``talk about puppies instead of ships'', is ``the nuclear way'': full reset, with the loss of all prior learning. \pSaab{} readily acknowledged this is unsatisfactory but offered it as an honest description of the current state.

Statistical trust is the {\em most demanding paradigm to operationalize} because it requires the surrounding organizational and assurance infrastructure to accept distributional success criteria. \pSaab{} himself observed that the customers of their systems, i.e.\ defence procurement organizations, ``struggle to define'' what acceptable distributional success looks like, even when they accept in principle that binary correctness is the wrong frame.

The statistical paradigm partially overlaps with the engineering paradigm: \pVCC{}'s ``forward-looking verification plans'' implicitly accept that single-point assurance is insufficient and that evidence must be collected across deployments. The difference is that the statistical paradigm pushes probabilistic acceptance into the definition of trustworthiness itself, whereas the engineering paradigm tries to constrain probabilistic behavior as much as possible.

\subsubsection{Containment-based trust}

Containment-based trust treats the AI model itself {\em as fundamentally untrusted, and locates trust in the surrounding architecture}: zero-trust isolation, retrieval-augmented grounding, data masking, the separation of generator and guardian, and the architectural separation of AI components from deterministic system cores.

\pNGDI{} (national government digital infrastructure), leading AI-enabled public digital services for a national government's flagship citizen-facing platform, articulates this paradigm explicitly. Operating in a context where his AI systems process the personal data for millions of citizens, \pNGDI{} adopts a zero-trust posture toward the model:

\begin{quote}
``We don't have policies related to the AI, so we're like a wild west, you know, so we can do everything we want, we don't have real legislation. But we have some, let's say, team maturity and
understanding of our data we are [dealing] with, so we put the rule that zero trust LLM, so LLM
never see real PII data [..] that's the first item.''
\end{quote}

Trust, in this paradigm, is not required of the model. The model never sees raw PII; data masking is mandatory and retrieval grounding constrains outputs to documented sources. The model is treated as a powerful but inherently untrustworthy component, and trustworthiness is engineered into the architecture surrounding it. \pNGDI{} also reports operating in a context of remarkably high adoption, where approximately 94\% of development staff use GenAI tools daily, but with governance practices explicitly racing to catch up. The containment paradigm is what allows the high adoption to be tolerated while assurance practices mature.

\pEricsson{}'s generator-guardian separation principle is the architectural sibling of \pNGDI{}'s zero-trust posture. The principle holds that the AI component generating outputs and the AI component checking those outputs should be independently developed, independently trained, and architecturally separated. The analogy he draws is to the long-standing software engineering principle that developers should not test their own code. Indirectly, the same principle surfaces in \pEduTS{}'s description of cross-model validation: outputs from one LLM are checked by a different LLM as a defense against the failure modes of any single model.

\pSaab{}'s hybrid architecture consisting of a deterministic data layer plus LLM interpretation layer is another instance of the containment paradigm, working in concert with his statistical-trust framing. The deterministic core preserves engineered guarantees; the LLM layer is contained to interpretation tasks where bounded variability is acceptable.

Several enterprise-services respondents adopt partial containment without naming it as such. \pFinAn{} describes guardrails and the use of enterprise AI services as a way of confining where AI runs. \pRetailBank{} describes his bank's use of RAG to ensure that customer-facing assistants answer ``based on the RAG and not anything random.'' The paradigm is not specific to safety-critical or public-sector contexts; it appears wherever sensitivity to model failure is high relative to the perceived intrinsic reliability of the model.

This paradigm's characteristic limitation is architectural cost. Containment is expensive as it requires duplicate models, infrastructural complexity, data segregation and explicit grounding pipelines. Organizations whose trust requirements are weaker pay less for containment infrastructure and accept the operational or engineering paradigms instead.

\subsubsection{Paradigms as orthogonal cross-cuts}
These four paradigms are not levels on a ladder. They are different epistemological strategies for reconciling the convergent operational definition of trustworthiness with the probabilistic reality of generative AI. They cross-cut the dimensions of the health check model we present in Section 5: an organization is characterized both by its position on each dimension and by the trust paradigms it operates under. A safety-critical engineering-trust organization (\pVCC{}'s automotive ADAS context) and an enterprise SaaS operational-trust organization (\pFlight{}'s aviation planning context) may score similarly on some dimensions while the meaning of those scores differs because the underlying epistemologies differ.

We return to the implications of this cross-cutting structure in Section \ref{sytemlayerdimensions} and to its consequences for the alignment hypothesis in Section \ref{alignment}.

\subsection{The dimensions along which organizations vary}
\label{dimvary}

\begin{figure}[htbp]
\includegraphics[width=\textwidth]{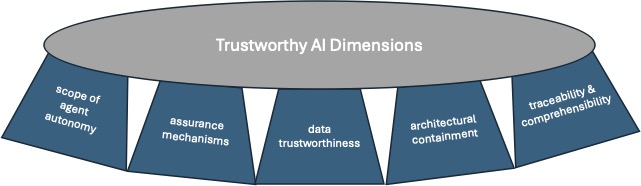}
\caption{Dimensions: System Layer}
\label{fig:systemlayer}
\end{figure}

Beyond the choice of paradigm, the corpus surfaces eight identifiable dimensions along which trustworthy autonomy varies in practice. We organize the dimensions into two layers: five describing properties of the {\em system} itself, and three describing properties of the {\em organization} in which the system is developed and operated. In figure \ref{fig:systemlayer}, the system layer dimensions are presented graphically. The dimensions are presented here in the order they appear in the health check model (Section \ref{TAHCM}); we discuss each briefly with its empirical anchoring before formalizing the levels in Section \ref{TAHCM}.

In the context of this paper, we use the words {\it dimension} and {\it paradigm} as part of the model that we inductively derive. We define these as follows: A {\bf dimension} is an axis of variation along which organizations differ in how they operationalize trustworthy autonomy, defined such that an organization's position on it can be characterized largely independently of its position on the others. Each dimension captures a distinct aspect of the system or organization that must be made trustworthy. This is opposed to a {\bf paradigm}, which captures the strategy by which trust is established. Each dimension is expressed as an ordered set of five qualitatively distinct levels; the levels are analytical categories describing positions along the axis, not stages of a maturity progression, so a higher level denotes greater scope or capability rather than a better or more desirable state.

\subsubsection{Scope of Agent Authority (system layer)}

The scope of agent authority concerns {\em what AI is allowed to do, across what parts of the software lifecycle, and at what abstraction level.} The corpus shows substantial variation. At one end, several safety-critical organizations deliberately limit AI to local code-completion tasks within human-defined components. At the other end, \pNGDI{}'s organization permits AI agents to participate in requirements generation, code generation, test generation, and even legal document analysis, with humans operating largely in oversight rather than direct-control roles.

Respondents consistently described authority as varying across the lifecycle within a single organization. \pEntMS{}'s organization, for instance, runs autonomous agents for legacy-system reverse engineering and modernization workflows, with high authority at the level of integration, while keeping human control firmly over high-stakes architectural decisions. \pEduTS{} describes the same pattern from the opposite direction: AI authority is widespread for coding and code review (LLMs reviewing other LLMs) but architecture remains a human domain.

A second pattern is the recursive structure of authority across architectural layers. \pVCC{}'s automotive context, even within ADAS, exhibits low AI authority at system-level architecture, moderate authority at component construction, and growing authority at module-level refactoring. \pSubsea{}'s AUV systems show similar layering. We return to this multi-level structure later in the paper.

\subsubsection{Assurance Mechanisms (system layer)}

Assurance concerns {\em how confidence in correctness, safety, and behavioral conformance is established}. The corpus is dense in this dimension; it is the operational core of trustworthy autonomy as practitioners actually encounter it.

Across the Kazman set of interviews in particular, evaluation emerged as the most named mechanism. \pMES{} (mixed enterprise services) framed evaluation as the single core mechanism for trustworthy AI systems, distinguishing it from superficial correctness by emphasizing that evaluation must include checks for whether agents are ``cheating'', for instance by hard-coding outputs to pass tests rather than actually solving the underlying problem. \pESS{} (enterprise software services) described evaluation as essential but undermined by benchmark contamination---the practical difficulty of constructing benchmarks the model has not seen during training. \pEntMS{} described ``quality gates'' and ``validation agents'' as the structural form evaluation takes in his organization.

In the safety-critical domains, assurance takes the form of formal frameworks. \pVCC{} is involved in the ISO PAS 8800 standardization effort, which extends data-management-lifecycle thinking into AI-relevant assurance for automotive contexts. The frameworks are visibly trying to catch up, rather than leading; \pVCC{}'s self-assessment is that the methodology lag is the central problem.

In the enterprise services domains, assurance is largely emergent and process-based. \pEduTS{}'s organization, mid-sized and actively integrating AI into the SDLC, describes the need for ``constant validation'' but has not yet systematized it. \pFlight{} describes assurance as relying on ``ground truth datasets'' and feedback loops; \pFinAn{} (financial analytics) describes treating AI ``as a junior developer'' and using multi-model cross-checking as a validation technique. Across the enterprise set, the consistent observation is that the validation burden falls on humans, and that this burden is growing faster than human review capacity.

\subsubsection{Data Trustworthiness (system layer)}

Data trustworthiness concerns {\em ownership, lineage, quality, exposure protection, and grounding}. The corpus is rather dense in this dimension. It surfaced in twelve of the eighteen interviews and appeared as the binding structural constraint in several.

Sovereignty and export control set hard constraints in \pSaab{}'s case. \pSaab{}'s AI infrastructure must physically reside in their home jurisdiction under national defence export-control regulations, which significantly limits the choice of available models and infrastructure. \pNGDI{}'s zero-trust posture around personally identifiable information is similarly structural; the architecture must guarantee that raw PII is never visible to the model.

Data ownership across multi-actor ecosystems was a recurring concern. \pSubsea{} described the AUV-data ownership problem in stark terms:

\begin{quote}
``You're using a sensor from Company A that is fitted onto an AUV from Company B, but the data is being managed by the operator of that AUV, which is Company C. But you're taking pictures or scans of the operator's field, so that's our data, so\ldots\ how are you going to? Who's going to own it? Who has rights to it?'' (\pSubsea{})
\end{quote}

He noted that every major energy operator has its own answer to these questions, and that the absence of consensus is itself a major constraint on how AI can be deployed.

\pFlight{} similarly described data ownership as a binding limit for his organization: ``We maybe don't own as much data as we as you would have wanted. There's a lot of customer-owned data and we don't have full access to it.'' \pFinSer{} identified privacy, data protection, and licensing as core elements of his definition of trustworthiness. \pRetailBank{} described legacy systems and ``data unlockability'' as the practical constraint that has forced his bank's AI work onto limited data subsets.

In the enterprise services contexts, data trustworthiness concerns shift toward IP and licensing: \pFinAn{} raised concerns about IP exposure and the risk of code leaks; multiple Kazman respondents described the legal ambiguity around training data and the divergence between US and EU regulatory expectations as binding constraints on what infrastructure can be used.

\subsubsection{Architectural Containment (system layer)}

Architectural containment concerns the separation of AI from deterministic system cores, the isolation of AI components from one another, and the construction of independent guardian systems. This dimension is implicit in the containment-based trust paradigm (§4.2.4) but applies more broadly.

\pEricsson{}'s generator-guardian separation principle is the clearest single articulation of this mechanism: the AI generating outputs and the AI checking those outputs should be independently developed and architecturally separated. In the words of \pEricsson{}: ``To develop both the test cases that validate your code as the one that develops the code each. Otherwise you test what you implemented and not what you should. Similar patterns happens. I think when we have this AI generated code and AI generated guardrails, these need to be separate.'' \pSaab{}'s hybrid architecture, consisting of a deterministic data layer plus a probabilistic LLM interpretation layer, is a different instantiation of the same principle.  \pNGDI{}'s zero-trust architecture, with mandatory data masking, is the most extensive instantiation in our corpus. Also \pWind{} independently stresses the importance of separating AI functionality from the rest of the system.

Less extreme forms of containment appear widely. \pEduTS{} describes cross-model validation as a pattern; \pRetailBank{} describes RAG grounding as the standard mechanism for confining customer-facing assistants. This dimension is the architectural counterpart of the assurance dimension: where assurance asks how we establish confidence, containment asks where AI is allowed to act and what surrounds it when it does.

\subsubsection{Traceability and Comprehensibility (system layer)}

Traceability concerns whether system decisions can be reconstructed and understood; comprehensibility concerns whether anyone still understands what the system as a whole does, particularly over time as AI-generated components accumulate.

{\em Traceability} emerges most strongly in regulated and public-sector contexts. \pRetailBank{} identified the lack of traceability as a ``significant issue'' at his organization, manifesting as shadow IT, with users copying AI-generated content into their workflows without traceable provenance. \pVCC{} described traceability as part of the data-management lifecycle that ISO PAS 8800 is intended to formalize. \pEduTS{} named traceability and explainability as future improvement priorities in his organization.

{\em Comprehensibility} is the concern that the system as a whole becomes opaque.  \pNGDI{} described the risk in terms of ``loss of architectural visibility'' and the resulting difficulty of debugging. \pFinAn{} described the same phenomenon at the level of generated code: hidden bugs in large codebases, where the volume and pace of AI generation outstrips human capacity to maintain understanding.  The ``vibe coding'' failure mode he named is exactly this: code that appears to work but that no one can validate, much less maintain.

We treat traceability and comprehensibility as the same dimension because the underlying organizational competence, knowing what the system does and being able to explain it, is shared. In the model we offer in Section 5, the dimension is named compactly as Traceability \& Comprehensibility, with the levels formalized to capture both decision-level reconstruction and system-level understanding.

\subsubsection{Governance (organizational layer)}

Governance concerns {\em policies, sanctioned use, role definitions and accountability structures} — the explicit organizational rules under which GenAI is permitted to be used and what its outputs can be used for.

Governance varies dramatically across the corpus. At the encoded end, \pRetailBank{}'s bank operates under an explicit policy mandate that human-in-the-loop is required for all AI systems; \pRetailBank{} described this as a corporate commitment that constrains every AI deployment. At the emergent end, multiple enterprise services organizations describe governance as evolving in parallel with adoption: \pEntMS{}'s organization has an internal ``center of excellence'' working on methodology; \pNGDI{}'s organization is racing to formalize policies in the wake of near-universal adoption (``Wild West externally, internal policies emerging'').

The Bosch--Olsson subset of interviews shows governance most clearly at the safety-critical end: \pSaab{}'s defence context is governed by national defence-export regulations that determine what infrastructure is permissible.  \pVCC{}'s automotive context is governed by automotive safety standards (ISO 26262, ISO PAS 8800) that determine what assurance evidence is required. In these contexts governance is largely external and set by regulators and standards bodies, with the organization adapting its practices to comply.

A consistent observation is that governance is the dimension where organizations differ most by sector. Enterprise services with no external regulator must invent governance internally; regulated industries inherit it from outside. This pattern shapes the domain-conditional analysis we present in Section \ref{domainpatterns}.

\subsubsection{Human Oversight Posture (organizational layer)}

Human oversight posture {\em concerns where humans remain in/on/out of the loop}. The corpus shows near-universal agreement that human-in-the-loop is the current default, but the trajectory toward human-on-the-loop (oversight rather than direct control) is widely anticipated.

\pRetailBank{}'s organizational mandate is the most explicit policy formulation. \pSubsea{}'s account is equally direct:

\begin{quote}
``I think humans need to always be in the loop at the start. [\ldots] For the energy industry, I wholeheartedly support human in the loop. There are some elements where you can have human on the loop. [\ldots] Once we train and we build up that confidence in the system, then of course we can see how we can incrementally release the human.'' (\pSubsea{})
\end{quote}

The same in-then-on trajectory is articulated by \pEduTS{}, by \pEntMS{}, and (in a different vocabulary) by \pVCC{}, who treats human review as the primary current trustworthiness mechanism but anticipates a future in which AI handles much of the routine assurance work, freeing humans for the design-of-product activities that, in his view, will remain irreducibly human:

\begin{quote}
``I think [humans are] definitely [needed] for the product definition that we understand. All related components. One is the product will be much more widened now. In the past we said, yeah, it's the system or the software is the product, but the product will be much bigger now. And there I think the [humans] will be much more informed in the definition of it, because you can rely on many more areas.''
\end{quote}

The defence-surveillance case (\pSaab{}) is unusual in that national law currently mandates human responsibility for lethal autonomous decisions, a hard constraint that prevents human-on-the-loop deployment regardless of technical capability. \pSaab{} was explicit that the legal constraint is the binding factor, not the technical one, and that in an active conflict the legal framework would change rapidly.

\subsubsection{Workforce Capability Sustainability (organizational layer)}

Workforce capability sustainability concerns the {\em preservation of organizational expertise, the development pipeline for junior practitioners, and the broader question of whether the organization retains the human capabilities} required to oversee, validate, and evolve AI-driven systems over time.

This dimension surfaced more prominently than we anticipated. \pFinAn{} articulated it most sharply, describing what he called a K-shaped productivity outcome: senior engineers, equipped with GenAI tools, become substantially more productive; junior engineers, lacking the deep system understanding needed to validate generated outputs, become less effective. The longer-term concern is the seed-corn problem --- that the production pipeline for the next generation of senior engineers may be undermined if junior engineers are sidelined by AI tools today.

\pRetailBank{} described the same phenomenon at his organization in terms of skill degradation: employees who outsource core thinking to AI lose the underlying capability and become unable to critically evaluate AI outputs. \pNGIT{} described it in terms of shallow solutions and bloated code: junior developers accepting AI suggestions without the experience to assess quality. \pESS{} described it as ``systemic knowledge loss.'' P11 observed that ``people-related challenges'' --- skills, mindset, adoption --- are the biggest issue in his organization, ahead of technical challenges. \pWebser{} expresses concerns about the fire-then-rehire-juniors pattern.

Workforce capability sustainability is the dimension most clearly organizational rather than technical. It cannot be addressed by tooling alone, and it interacts with the assurance dimension: the validation burden that several respondents named as the central technical challenge will fall on a workforce whose capacity to bear it depends on how the organization manages the development of expertise. Several respondents --- notably \pNGDI{}, whose organization is creating ``AI trainer'' roles, and \pMES{}, who described emerging roles like ``intelligence engineer'' and ``intelligence architect'' --- describe explicit attempts to design new workforce roles in response to this concern.

\begin{figure}[htbp]
\includegraphics[width=\textwidth]{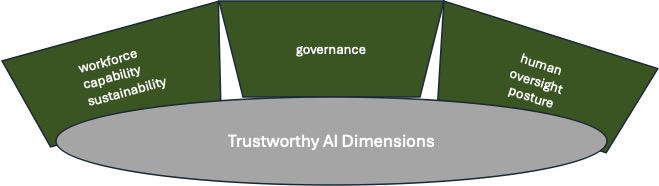}
\caption{Dimensions: Organizational Layer}
\label{fig:orglayer}
\end{figure}

In figure \ref{fig:orglayer}, we present the three dimensions of the organizational layer.

\subsection{Domain-specific patterns}
\label{domainpatterns}

The corpus exhibits three clearly distinguishable patterns across domain contexts. We summarize them here as a precursor to the configuration analysis in Section \ref{threats}.

{\bf Safety-critical and regulated}: Organizations in this group (\pEricsson{}/telecom, \pSaab{}/defence surveillance, \pSubsea{}/subsea AUV-energy, \pVCC{}/automotive ADAS, \pRetailBank{}/re\-tail banking, \pNGDI{}/national government infrastructure in part) exhibit dominantly engineering or containment paradigms, with assurance as the binding constraint and oversight either mandated by law or required by industry policy. Agent authority is thus deliberately limited. Data trustworthiness is heavily constrained by sovereignty, export controls, or sensitivity regimes. Governance is largely external --- set by regulators or standards bodies. Workforce capability is managed but rarely framed as a primary risk.

{\bf Enterprise software services}: Organizations in this group (\pMES{}, \pESS{}, \pEduTS{}, \pEntMS{}, \pFinSer{}, \pWebser{}, \pInsur{}, \pFinAn{}) exhibit mixed paradigms, often combining operational trust with emerging engineering practices. Agent authority is high in development workflows, especially in coding and documentation, but generally low in product runtime. Governance is internal and emergent rather than externally imposed. The binding concerns are the validation burden and workforce capability sustainability --- the K-shape problem looms larger here than in the safety-critical group, both because the workforce is more directly exposed to AI in daily work and because the regulatory floor that protects assurance in safety-critical contexts is absent.

{\bf Exploratory and transformation contexts}: Organizations in this group (e.g. \pNGDI{}, \pNGIT{} and \pCongl{}) exhibit the least developed trust paradigm, with agent authority being negotiated organizationally and structural questions about data and governance dominating technical ones. The pattern is most visible at the early phase of adoption, when the question is less ``how do we ensure trustworthiness'' than ``what should the organization do with this technology at all.''

The three patterns are not exhaustive, nor are they exclusive — many real organizations exhibit a mix — but they capture the dominant tendencies in our sample. Section \ref{TAHCM} elaborates them through specific configurations.

\subsection{Recurring tensions}
\label{tensions}

Several tensions surfaced across the corpus that the dimensional structure alone does not fully capture. Below, we detail four of these tensions:

{\bf Speed versus assurance}: \pVCC{}'s ``we don't have trustworthiness at all'' is the sharpest expression: the speed of GenAI iteration outpaces the construction of assurance methodologies designed for slower, more discrete release cycles. The tension is shared by \pSaab{}, \pRetailBank{}, \pCongl{} and several enterprise services respondents who describe AI generation as fast and AI validation as slow. This asymmetry creates the conditions under which the alignment hypothesis (discussed in section \ref{trustparadigms}) can be tested: organizations whose authority outruns their assurance face pathological risk pressure. Several interviewees, including \pWebser{}, \pCongl{} and \pWind{}, stress that hallucination by AI is inevitable and that we need mechanisms to catch those hallucinations when they happen.

{\bf AI for development versus AI in products}: Several respondents (\pEricsson{}, \pFlight{}, \pSubsea{}) explicitly distinguish the two cases, recognizing that the trustworthiness concerns are different. But the dimensional structure of trust does not always cleanly distinguish them in practice, many of the same dimensions apply to both, while the levels of concern differ. \pFlight{} captured the issue directly when he asked the interviewers, in the opening of his interview, which kind of AI use the questions were about; he resisted answering trustworthiness questions in the abstract without that disambiguation.

{\bf Continuity versus discontinuity}: \pFlight{} and \pRetailBank{} normalize: GenAI is, in their framing, an acceleration of pre-existing optimization opacity rather than a categorical break. \pSubsea{} and \pVCC{} problematize: GenAI is a new risk requiring methodological extension. \pSaab{} goes furthest in the discontinuity direction: he questions whether software, as a fixed artifact, even survives as a concept (``do we need software in the way we have software now?''), as agentic systems generate code on the fly in response to intent rather than against fixed specifications. The continuity-versus-discontinuity reading shapes which paradigm an organization adopts.

{\bf Productivity versus comprehensibility}: \pFinAn{}, \pNGDI{}, \pNGIT{}, \pWind{} and others surface a consistent observation: GenAI generates more code, faster, but no one is sure they fully understand it. The volume of generation is outpacing the human capacity to comprehend, validate, and maintain. The K-shape problem is a downstream consequence; the loss of architectural visibility is another; the rise of ``vibe coding'' as a recognized failure mode is a third. The tension is most acute where AI authority is highest and where traceability and comprehensibility are weakest --- exactly the alignment problem we formalize in Section \ref{alignment}.

These tensions are not artifacts of the inductive coding; they are observable patterns in the data that any account of trustworthy autonomy in GenAI-assisted software engineering must confront.

\section{The Trustworthy Autonomy Health Check Model}
\label{TAHCM}

Section \ref{findings} presented the four trust paradigms and the eight dimensions as inductively emergent findings from the interview corpus. In this section we formalize those findings into the Trustworthy Autonomy Health Check Model. The model is offered as an analytical instrument for characterizing how organizations operationalize trustworthy autonomy in software-intensive systems incorporating generative AI. It is not a maturity ladder; it is a structured configuration space.

We proceed in five steps. Section 5.1 presents the overall structure. Section 5.2 formalizes the eight dimensions, each having a five-level ordinal scale. Section 5.3 explains how the four trust paradigms cross-cut the dimensions. Section 5.4 states the alignment hypothesis — the model's central testable claim — and the empirical support for it. Section 5.5 names what the model deliberately does not do.

\subsection{Overview of the Model}

The Trustworthy Autonomy Health Check Model is organized into two layers and a cross-cutting paradigm overlay.

The {\bf system layer} comprises five dimensions describing properties of the software-intensive system itself: Scope of Agent Authority, Assurance Mechanisms, Data Trustworthiness, Architectural Containment, and Traceability \& Comprehensibility. These are the technical and architectural properties that, taken together, determine how AI is permitted to operate within the system and how its behavior is constrained, validated, and rendered understandable.

The {\bf organizational layer} comprises three dimensions describing properties of the organization in which the system is developed and operated: Governance, Human Oversight Posture, and Workforce Capability Sustainability. These are the social-organizational properties that determine how the human organization governs, supervises, and sustains its capacity to engage with AI-driven systems over time.

The paradigm overlay is used to identify the dominant trust paradigms under which the organization operates: operational, engineering, statistical, or containment-based. Each paradigm describes how trust may be established and the dimensions describe what aspects of the system and organization need trustworthiness. Together they characterize an organization's position in the configuration space.

%Earlier in the paper, we presented the model schematically. The eight dimensions are divided between five system-layer dimensions  three organizational-layer dimensions. The four trust paradigms appear as cross-cutting, indicating that paradigm choice characterizes the way trust is established across all dimensions rather than constituting a ninth dimension.

%\begin{figure}[htbp]
%\includegraphics[width=\textwidth]{TAI_overview.jpg}
%\caption{The Trustworthy Autonomy Health Check Model.}
%\label{fig:health-check-model}
%\end{figure}

%\textcolor{red}{Figure needs to be replaced}

A critical interpretive point: {\bf\em the model is not intended to produce a single aggregated score}. Each dimension carries its own analytical weight; collapsing them to a single number would reintroduce the unidimensional compression we have spent the paper arguing against. Healthy configurations are those in which the dimensions are aligned, not those in which all dimensions reach the highest level (we will return to this point in  Section \ref{trustparadigmscross}). Different domains require different alignments; the model supports the diagnosis of misalignment, not the prescription of a single target.

\subsection{The Eight Dimensions Formalized}

For each of the eight dimensions we define a five-level ordinal scale. The levels are analytical categories describing qualitatively distinct positions along the dimension, not stages of progression. A higher level is not inherently better than a lower one; rather, {\em higher levels imply greater capability} (or greater scope, depending on the dimension) that may or may not be appropriate for the context. The five-level structure is consistent across dimensions to facilitate comparison, but the meaning of each level is dimension-specific.

For each of the dimensions, we present the levels in compact tabular form below.

\subsubsection{System-Layer Dimensions}
\label{sytemlayerdimensions}

In Table \ref{tab:system-dimensions}, we present the five system-layer dimensions and their five levels. Levels are analytical categories, not maturity stages: higher levels are not inherently better and may be inappropriate for low-stakes contexts.

\begin{table*}[htbp]
\centering
\caption{The five system-layer dimensions and their five levels.}
\label{tab:system-dimensions}
\footnotesize
\renewcommand{\arraystretch}{1.25}
\begin{tabularx}{\textwidth}{@{}p{1.8cm} *{5}{>{\raggedright\arraybackslash}X}@{}}
\toprule
\textbf{Dimension} & \textbf{L1} & \textbf{L2} & \textbf{L3} & \textbf{L4} & \textbf{L5} \\
\midrule
\textbf{Scope of Agent Authority} &
Assistive: AI assists local tasks under direct human control. &
Workflow-bounded: AI operates within defined workflows (CI/CD, configuration). &
Component-bounded: AI acts within delineated subsystems with explicit interfaces. &
System-bounded: AI participates in system-level work within encoded constraints. &
Cross-system: AI operates across system and organizational boundaries with negotiated authority. \\
\addlinespace[2pt]
\textbf{Assurance Mechanisms} &
Ad hoc: trust rests on individual judgment; no systematic validation. &
Manual review: human review at defined gates; informal evaluation. &
Structured: codified review processes, test gates, evaluation frameworks. &
Continuous: automated validation in tooling; runtime monitoring; evidence generation. &
Adaptive: assurance mechanisms evolve from operational data; forward-looking verification across multiple releases. \\
\addlinespace[2pt]
\textbf{Data Trustworthiness} &
Opportunistic: data is used as available; ownership and lineage unaddressed. &
Documented: ownership and access are known but not enforced; quality variable. &
Controlled: documented lineage, access controls, basic quality checks. &
Constrained: enforced sovereignty / sensitivity controls; mandatory grounding for sensitive operations. &
Architecturally guaranteed: data exposure prevented by design; zero-trust around sensitive data; masking and segregation enforced. \\
\addlinespace[2pt]
\textbf{Architectural Containment} &
None: AI integrated throughout the system without explicit isolation. &
Wrapped: AI components have explicit interfaces but share trust boundary with deterministic code. &
Bounded: AI components isolated to specific subsystems; outputs validated at boundaries. &
Hybrid-separated: deterministic core preserves engineered guarantees; AI confined to interpretation or augmentation layer. &
Zero-trust: AI never trusted; independent guardian systems, generator-guardian separation, redundant cross-validation. \\
\addlinespace[2pt]
\textbf{Traceability \& Comprehensibility} &
Opaque: AI decisions cannot be reconstructed; system behavior accumulates without explanation. &
Post-hoc: explanations available on demand but not systematically retained. &
Logged: decision traces automatically recorded and queryable. &
Linked: traces connected to requirements, architectural constraints, and policies. &
End-to-end: full traceability across abstraction levels; system-level comprehensibility maintained as AI generation scales. \\
\bottomrule
\end{tabularx}
\end{table*}

A few interpretive notes on these dimensions:

\begin{itemize}
\item For {\it Scope of Agent Authority}, level 5 (``Cross-system'') was articulated in the corpus only as an aspirational direction (most explicitly in \pFlight{}'s vision of an autonomous airline-management system and \pSaab{}'s intent-based agentic platform vision). No respondent described their current organization as operating at level 5. We include it because the corpus consistently points toward it as the desired destination.

\item For {\it Assurance Mechanisms}, the gap between level 3 (``Structured'') and level 4 (``Continuous'') is the largest  gap in the corpus. Most enterprise services respondents operate at level 2 or 3 with aspirations toward level 4; \pVCC{}'s automotive context targets level 5 but reports level 3 as their actual capability, which is the central tension he articulates.

\item For {\it Data Trustworthiness}, the spread is unusually wide. Several enterprise services organizations remain at level 1 or 2 even as they deploy AI broadly; safety-critical and PII-sensitive contexts (\pSaab{}/defence, \pNGDI{}/national government infrastructure, \pRetailBank{}/retail banking) operate at level 4 or 5 by necessity.

\item For {\it Architectural Containment}, level 5 is the explicit endpoint of the con\-tainment-based trust paradigm. \pNGDI{}'s zero-trust architecture and \pEricsson{}'s gene\-rator-guardian separation principle are the corpus's clearest level-5 instantiations.

\item For {\it Traceability \& Comprehensibility}, level 5 is currently aspirational across the entire corpus. Several respondents (\pFinAn{}, \pNGDI{}, \pRetailBank{}) articulated the failure of low traceability under high AI authority more vividly than any respondent articulated the achievement of high traceability.
\end{itemize}

\subsubsection{Organization-Layer Dimensions}

In Table \ref{tab:org-dimensions}, we present the organizational-layer dimensions and their five levels. As with the system-layer dimensions in Table~\ref{tab:system-dimensions}, levels are analytical categories rather than maturity stages.

\begin{table*}[htbp]
\centering
\caption{Organizational-layer dimensions and their five levels. }
\label{tab:org-dimensions}
\footnotesize
\renewcommand{\arraystretch}{1.25}
\begin{tabularx}{\textwidth}{@{}p{1.8cm} *{5}{>{\raggedright\arraybackslash}X}@{}}
\toprule
\textbf{Dimension} & \textbf{L1} & \textbf{L2} & \textbf{L3} & \textbf{L4} & \textbf{L5} \\
\midrule
\textbf{Governance} &
Absent: no explicit policies; use is opportunistic. &
Emerging: guidelines exist but are not enforced; sanctioned tools partially defined. &
Documented: explicit policies, sanctioned tools, role definitions. &
Enforced: policies operationalized in tooling and processes; accountability structures defined. &
Adaptive: policies evolve from operational evidence; integrated with regulatory frameworks. \\
\addlinespace[2pt]
\textbf{Human Oversight Posture} &
None: AI outputs accepted without systematic human review. &
Reactive: humans intervene after problems are detected. &
Human-in-the-loop: humans review AI outputs at defined points before action. &
Human-on-the-loop: humans monitor and intervene exceptionally; calibrated to action risk. &
Strategic: humans define intent, constraints, and exception protocols; AI operates within these with limited  supervision. \\
\addlinespace[2pt]
\textbf{Workforce Capability Sustainability} &
Eroding: junior pipeline weakening; expertise concentrating in seniors without replacement. &
Unmanaged: skill effects of AI not systematically addressed; risks acknowledged but not mitigated. &
Monitored: skill effects tracked; some training and oversight in place. &
Managed: deliberate training pipelines, role redesign, validation skills development. &
Adaptive: organizational learning embedded; new roles institutionalized; expertise renewal mechanisms operational. \\
\bottomrule
\end{tabularx}
\end{table*}

We offer a number of observations regarding these organizational dimensions:

\begin{itemize}
\item For {\it Governance}, the corpus shows the widest empirical variation. \pRetailBank{}'s mandatory human-in-the-loop policy is the clearest level 4 instantiation. Most enterprise services respondents operate at level 2 (emerging guidelines, internal centers of excellence) or level 3 (documented policies, sanctioned tooling). \pVCC{}'s safety-critical context is the only one in the corpus that explicitly targets level 5, through engagement with external standards development (ISO PAS 8800).

\item For {\it Human Oversight Posture}, level 3 (human-in-the-loop) is the universal current default and is mandated by external constraint in several cases (national defence law for \pSaab{}; organizational policy for \pRetailBank{}). The trajectory toward level 4 (human-on-the-loop) is widely anticipated but currently exceptional. Level 5 was not described as currently operational by any respondent.

\item For {\it Workforce Capability Sustainability}, the corpus is conspicuously bimodal: most respondents describe their organization as level 1 or 2 (eroding or unmanaged), while a few (\pNGDI{} with AI-trainer roles, \pMES{} with new hybrid roles, \pRetailBank{} with explicit skill-degradation monitoring) describe deliberate level-3 efforts. No respondent described their organization at level 4 or 5; we include those levels as the structural endpoint suggested by the corpus's repeated articulation of the problem.
\end{itemize}

A critical caveat applies across all the levels: the empirical anchoring is necessarily partial. Several level definitions, especially at the upper end of each dimension, are extrapolations from how respondents articulated the direction of needed change rather than from observations of organizations currently operating at that level. 

\subsection{The trust paradigms as cross-cutting}
\label{trustparadigmscross}

\hyphenation{con-tain-ment}

The four trust paradigms — operational, engineering, statistical, and con\-tain\-ment\--based — cross-cut the dimensional structure rather than constituting a ninth dimension. They describe how an organization establishes trust; the dimensions describe what aspects of the system and organization need trustworthiness.

An organization is characterized, in the model, by (a) its position on each of the eight dimensions and (b) the trust paradigm or paradigms it dominantly operates under. The combination is more informative than either alone.

Consider two organizations at apparently similar positions on the system-layer dimensions: \pVCC{}'s automotive ADAS context (Scope of Agent Authority L2; Assurance L3 targeting L5; Data Trustworthiness L4; Architectural Containment L3; Traceability L3) and \pFlight{}'s aviation crew-planning context (Scope L2-3; Assurance L3; Data L3; Containment L2; Traceability L3). The positions are broadly similar. But the meaning of those positions differs sharply because \pVCC{} operates under an engineering trust paradigm (assurance is the binding constraint; failures must be vanishingly rare) while \pFlight{} operates under an operational trust paradigm (outcomes are continuously measured; failures are detectable and recoverable in production).

A dimension-only reading of these two organizations would obscure their fundamentally different theories of correctness. A paradigm-only reading would obscure the configurational variation within each paradigm --- for example, the difference between \pSaab{} and \pVCC{}, both engineering-trust organizations but with different containment postures and very different governance regimes.

The cross-cutting structure has a further consequence for the alignment hypothesis we present next. The same level on the same dimension may be sufficient under one paradigm and insufficient under another. Level 3 Assurance is sufficient under the operational paradigm when outcome measurement is reliable; it is insufficient under the engineering paradigm when failures cannot be tolerated; it is differently inadequate under the statistical paradigm when distributional success criteria require continuous evidence generation; it is irrelevant under the containment paradigm when the AI is not trusted to be assured in the first place.

\subsection{The alignment hypothesis}
\label{alignment}

The model's central testable claim is what we call the alignment hypothesis. We state it formally before unpacking it:

\begin{quote}
Effective trustworthiness is constrained by the alignment between the scope of agent authority and the enabling dimensions {\it Assurance Mechanisms, Data Trustworthiness, Architectural Containment, Human Oversight Posture}, conditional on the dominant trust paradigm. Misalignment produces either pathological risk (authority exceeds enabling) or unnecessary friction (enabling exceeds authority).
\end{quote}

The hypothesis decomposes the eight dimensions into two groups: {\it Scope of Agent Authority} (the dimension that defines what AI is permitted to do) and the enabling dimensions (the dimensions that determine whether the organization can responsibly support that authority). The remaining three dimensions — Traceability \& Comprehensibility, Governance, and Workforce Capability Sustainability — act as contextual qualifiers rather than first-order constraints in the hypothesis as stated; they influence the meaning of misalignment but are not its primary determinants.

The hypothesis has two failure modes. {\bf Pathological risk} arises when Scope of Agent Authority exceeds the enabling dimensions. The organization has granted AI more authority than its assurance, data trustworthiness, architectural containment, or human oversight can responsibly support. The result is an organization that is operating at a level of AI capability it cannot validate, contain, or supervise. Several respondents describe this configuration directly. \pEntMS{} is most explicit: high AI authority across the SDLC combined with what he frames as low trust in AI outputs and an explicit acknowledgement that the validation methodologies are not yet mature. \pNGDI{} describes the same pattern at a different scale: approximately 94\% adoption of GenAI in his organization with governance practices still under construction --- a configuration he characterizes himself as ``Wild West externally'' with internal policies racing to catch up. \pFinAn{} articulates the consequence at the level of workforce: senior engineers using AI effectively while juniors lose the capacity to validate, producing what he calls a K-shape that, over time, undermines the organization's capacity to support its own AI authority.

{\bf Unnecessary friction} arises when the enabling dimensions exceed the scope of authority. The organization has invested in assurance, data trustworthiness, containment, and oversight beyond what is needed for the AI it actually deploys. The result is an organization that pays for capability it does not use. \pRetailBank{}'s regulated banking context is a partial instance: strong governance (mandatory human-in-the-loop) and significant containment infrastructure (RAG grounding, sanctioned tooling) combined with limited AI authority (assistive use, no autonomous action) produces a configuration in which the enabling capability constrains the realization of value. \pRetailBank{} himself acknowledged the drag; he framed it as a deliberate trade in favor of risk reduction in a regulated context.

The hypothesis is conditional on paradigm. Under the operational paradigm, the binding enabling dimension is typically Assurance Mechanisms (because trust is established through outcome measurement and feedback loops); shortfalls in Architectural Containment are tolerable so long as outcomes can be observed. Under the engineering paradigm, all four enabling dimensions matter, but Assurance dominates. Under the statistical paradigm, Architectural Containment becomes critical (because containment is what makes statistical acceptance of variability tractable), and Assurance Mechanisms shift toward continuous and adaptive forms. Under the containment paradigm, Architectural Containment is the primary enabling dimension by definition; shortfalls elsewhere may be tolerable if the model is sufficiently contained.

This hypothesis is offered as a testable proposition rather than as an established finding. The corpus supports it but can not prove it. What we can claim is that the configurations that we have observed are consistent with it. Quantitative testing — operationalizing each dimension into measurable indicators, computing alignment metrics, correlating with downstream outcomes such as incident rates or productivity — is future work that we discuss in Section \ref{conclusion}.

What the hypothesis does offer right away is a diagnostic lens. An organization applying the health check is led to ask: where is my authority relative to my enabling dimensions? Where are my enabling dimensions relative to my authority? Under what paradigm am I (perhaps implicitly) operating, and is the configuration aligned with that paradigm? These are practical questions that the model is structured to answer.

\subsection{What the model does not do}
\label{doesnotdo}

To forestall over-extension, we name several things the model deliberately does not do.

\begin{itemize}
\item First, the model does not produce a single aggregated score. Combining the eight dimensions into one number would reintroduce the unidimensional compression we have argued against. Practitioners using the model are expected to engage with the configuration as a whole, not with a summary statistic.

\item Second, the model does not prescribe a target configuration. The healthy configuration depends on the domain, the system criticality, and the organization's strategic intent. A safety-critical organization may need to operate at level 5 on Assurance and level 2 on Scope of Agent Authority; an exploratory startup may rationally operate at level 1 on Assurance and level 4 on Scope. The model identifies misalignment given the organization's chosen direction; it does not choose the direction.

\item Third, the model does not predict outcomes. It is a structural and interpretive instrument. The alignment hypothesis is a testable claim about the relationship between configuration and outcome, but the hypothesis itself is not yet established. The model characterizes configurations; testing whether configurational alignment predicts incidents, productivity, or other outcomes is the natural next step in this research.

\item Fourth, the model does not address all dimensions of AI trustworthiness in the broad ethical sense. Notably absent are dimensions concerning fairness, bias, equity, and societal impact. These dimensions did not surface as primary concerns in our interview corpus — itself a finding worth noting. We do not claim that they are unimportant; we claim that they belong to a different empirical and conceptual stream that the present model does not attempt to incorporate.

\item Fifth, the model does not distinguish strongly between AI-for-development and AI-in-product. Several respondents (\pEricsson{}, \pFlight{}, \pSubsea{}) made this distinction explicit, and the dimensions in some cases manifest differently across the two cases. We treat the model as applicable to both, but practitioners should explicitly specify which case they are assessing when applying the model; an organization may have very different configurations for development-time and runtime AI even within the same system.

\end{itemize}

\section{Threats to Validity}
\label{threats}

We address threats to validity under the standard four-fold framing for qualitative research \cite{lincoln1985naturalistic}, \cite{maxwell2012qualitative}, adapted to the multi-team character of the present study. Several of these threats follow directly from methodological choices already disclosed in Section 3; we name them here without re-arguing the rationale.

\subsection{Construct validity}

Three threats to construct validity warrant specific attention.

{\bf Protocol use heterogeneity}: The most significant construct-validity concern in this study is that the three research teams applied the common interview protocol in slightly different ways. The constructs about which respondents spoke were inflected by the protocol under which they were interviewed. 

We mitigate this threat in two ways. First, the dimensional structure reported in Section 5 was tested for empirical support across the entire corpus and retained only where evidence appeared in at least two. The trust paradigms reported in Section 4.2 emerged most clearly in the Bosch–Olsson set but were corroborated by interviewees in both of the other sets. Second, we report the construct in the practitioner's own terms wherever possible, using verbatim quotation, rather than translating into a researcher-imposed vocabulary.

We acknowledge that the construct of "trustworthiness" was elicited rather than supplied. Respondents converged on an operational definition (Section 4.1) but may not have meant the same thing in fine-grained terms. We do not claim that ``trustworthiness" is a single well-defined construct; we claim that the {\it operationalizations} of trustworthiness across our respondents are sufficiently consistent and structured to support the dimensional model we report.

{\bf LLM-assisted summarization}: As disclosed in Section 3.4, all three research teams used generative AI tools to assist in producing per-interview summaries and codings. We address this transparently, both because the topic of the study makes silence on the matter untenable and because the practice introduces specific risks: paraphrasing drift, sub-claim aggregation, and loss of verbatim grounding. We mitigate these risks by retaining raw transcripts as the authoritative source, by treating LLM-generated summaries as candidate codings to be validated rather than as findings, and by drawing all verbatim quotations reported in this paper directly from the underlying transcripts rather than from AI-generated summaries.

{\bf Researcher prior commitments}: Some of the authors in the team earlier developed unidimensional maturity models instead of multi-dimensional model presented in this paper. This prior commitment shaped what we noticed in the data, the vocabulary in which we expressed findings, and the conceptual moves we considered. We acknowledge this openly in Section 3.6. The principal mitigation is the willingness, demonstrated in the paper itself, to revise the prior framing in light of empirical data; we treat the move from the unidimensional ladder to the multidimensional health check as evidence that the inductive process operated against rather than within our prior commitments.

\subsection{Internal validity}

Inductive interpretation involves judgment. We cannot prove that the eight dimensions are {\bf the} correct decomposition of trustworthy autonomy or that the four trust paradigms are exhaustive. We have argued for both decompositions from the data, but a different research team applying inductive analysis to the same corpus might reasonably produce a different decomposition.

Our cross-case synthesis may have over-weighted the most articulate or analytically distinctive respondents, such as \pSaab{}'s articulation of statistical trust, \pVCC{}'s framing of forward-looking verification, \pEricsson{}'s generator-guardian principle, and \pSubsea{}'s ``250 constituent parts''. These articulations are conceptually rich contributions that occupy substantial space in Section 4. The model is shaped by these voices to a degree that may exceed their statistical representation in the corpus. We do not regard this as a defect of the analysis --- distinctive cases often carry the conceptual load in qualitative work \cite{eisenhardt1989building}, \cite{yin2018casestudy} --- but we acknowledge that an alternative reading more heavily weighted toward typical cases might produce a different emphasis.

%\textcolor{red}{Second, the corpus contains several respondents whose accounts were captured in less analytical depth than others (the Muccini set in particular relied more heavily on AI-assisted summarization than the other two sets). We have tried to use these respondents as corroborating rather than as primary sources for the dimensional structure, but we acknowledge that the unevenness of analytical depth across the corpus is a source of internal-validity risk.}

\subsection{External validity / transferability}

We do not claim generalizability in the statistical sense. We claim transferability of the dimensional structure and the four paradigms to contexts similar to those represented in our sample, supported by the explicit reporting of participant context and of the configurations from which our findings were derived.

Several specific limitations of the sample bear on transferability. The sample skews toward Northern European, North American, and Eastern European software services contexts. Asian, Latin American, and African industry are represented only marginally (and that marginality is itself a limitation, since the construction of trustworthy AI is centrally bound up with regulatory environments that vary substantially across these regions). One interviewee from outside the dominant geographies (\pSubsea{}), with operational context spanning South Asia and the energy sector) provides partial corrective coverage; future work should systematically include voices from underrepresented geographies.

The sample skews toward larger and more sophisticated organizations. Small and earliest-adopter startups — which may exhibit the most extreme misalignments under the framework we propose — are underrepresented. Several Kazman respondents work at substantial (50.000+ FTE) software-services companies.  Our exploratory-context category in Section 4.4 is the most weakly empirically anchored of the three patterns and would benefit from purposive sampling of smaller, more experimental organizations.

The sample skews toward technical and strategic-technical voices: software architects, technical leads, AI strategy leads, and heads of engineering practice. We have less voice from compliance, legal, operations, and the lines of business consuming AI-assisted output. The governance dimension in particular would benefit from more compliance and legal perspectives; the workforce-capability dimension would benefit from HR and learning-and-development perspectives.

Finally, the sample skews toward respondents who were willing and able to articulate their organization's trustworthy-autonomy practices in a research interview. Organizations with the most opaque or least systematic practices may be the ones least likely to participate in research of this kind. The selection bias toward articulate respondents is a structural limitation of a voluntary interview-based research.

\subsection{Reliability / dependability}

Three independent coding teams produced converging interpretations of the corpus despite operating under different protocols and analytical conventions. We regard this convergence as one of the stronger forms of evidence available in inductive qualitative work — areas where all three teams independently identified the same theme carry stronger warrant than areas where only one team's data spoke to the theme. Section 4 marks (through explicit attribution) which themes are supported by which interviewer's set.

We retained peer-interview structured codings produced by each team as an audit trail. These would allow an independent researcher to reconstruct the path from raw transcripts through individual codings to the cross-corpus synthesis reported here, although we acknowledge that the LLM-assisted summarization step introduces variability that may be difficult to reproduce exactly.

We do not report a formal inter-coder reliability coefficient. The rationale for this choice was given in Section 3.4: such coefficients are contested in inductive qualitative work and would in any case be of limited interpretive value in a study where the three teams used slightly different protocols and operated in different stages of analysis. Reviewers who prefer formal reliability statistics may take this as a residual threat; we judge it to be a lesser threat compared to the methodological distortion that would have been required to manufacture an inter-coder reliability coefficient across interviews based on a protocol that was not designed to support one.

Replication of the inductive analysis by independent researchers would not be expected to produce identical results — inductive interpretation involves judgment — but should produce convergent themes.

\section{Conclusion and Future Work}
\label{conclusion}

This paper has reported an inductive interview study of how organizations across diverse industrial contexts operationalize trustworthy autonomy in soft\-ware-intensive systems incorporating generative AI. From eighteen interviews spanning safety-critical, public sector, regulated finance, and enterprise services domains, we have derived two principal findings.

First, organizations converge on what trustworthiness means — predictability, repeatability, and the absence of unintended behavior — but diverge sharply on how trustworthiness is established. We have identified four distinct trust paradigms: operational (trust as outcome measurement and feedback), engineering (trust as process discipline and staged validation), statistical (trust as bounded stochastic acceptance with runtime convergence), and containment-based (trust as zero-trust architectural isolation around an inherently untrusted model). Each paradigm is rationally adapted to a particular configuration of system, domain, and organization; organizations frequently exhibit several simultaneously across different subsystems.

Second, we have identified eight dimensions along which trustworthy autonomy varies in practice — five describing properties of the system itself (Scope of Agent Authority, Assurance Mechanisms, Data Trustworthiness, Architectural Containment, Traceability \& Comprehensibility) and three describing properties of the organization (Governance, Human Oversight Posture, Workforce Capability Sustainability). These dimensions are inductively derived; each is empirically anchored in eight or more of our interviews. The dimensions and paradigms together constitute the Trustworthy Autonomy Health Check Model: a structured configuration space that supports cross-organizational comparison without imposing a single progression path.

The model's central testable claim is the \textit{alignment hypothesis}: effective trustworthiness is constrained by the alignment between an organization's scope of agent authority and the enabling dimensions that determine whether it can responsibly support that authority, conditional on the dominant trust paradigm. Misalignment produces either pathological risk (authority exceeds enabling) or unnecessary friction (enabling exceeds authority). This hypothesis is offered as a proposition for quantitative future work to test rather than as an established finding.

The empirically grounded configurations presented in Section \ref{TAHCM} illustrate how the model is applied in practice. Each configuration corresponds to one or more real organizations in our corpus and shows that the same nominal dimensional position can be healthy in one domain and pathological in another. 

Four research directions extend the present work most naturally.

{\bf Quantitative testing of the alignment hypothesis}: The most immediate extension is to operationalize each of the eight dimensions into measurable indicators, define a misalignment metric (or set of metrics) over those indicators, and correlate misalignment with downstream outcomes such as incident rates, defect density, productivity, and time-to-detection. Such work would require either large-scale survey instruments distributed across many organizations or detailed longitudinal case studies in a smaller number of organizations. Both approaches face the standard challenges of measuring qualitative properties through quantitative proxies, but the structural hypothesis is sharp enough to support either.

{\bf Longitudinal study of dimensional evolution}: The present paper is a cross-sectional snapshot. A natural complementary study would track organizations over time to observe how dimensional configurations evolve — in particular, whether organizations move toward greater alignment as they mature in their GenAI adoption or whether misalignment is a persistent structural feature. Several of our respondents articulated explicit trajectories ("we are at level X, moving toward level Y"); whether such trajectories are actually realized is an empirical question that cross-sectional research cannot answer.

{\bf Recursive and multi-level health checks}: Section \ref{TAHCM} noted that within a single organization, dimensional configurations may differ substantially across architectural layers — system, subsystem, component, module. A natural extension of the present work is to develop a multi-level application of the health check: assessing each architectural layer separately and examining whether the alignment hypothesis holds at each layer or whether layer-specific patterns emerge. This would also engage productively with the recent literature on software architecture in the age of generative AI.

{\bf Domain-specific refinement}: The present paper's dimensions and pa\-ra\-digms are intended to be cross-domain. Some domains may, however, warrant more fine-grained dimensions. Safety-critical contexts may require splitting Assurance Mechanisms into design-time and runtime assurance; financial-regulatory contexts may require splitting Governance into internal-policy and external-regulatory governance; public-sector contexts may require explicit treatment of citizen-facing trustworthiness as a distinct dimension. Domain-specific elaborations of the model would test whether the cross-domain structure holds up under more granular treatment.

Several other research directions are valuable but belong to a normative research stream that builds on rather than directly extends the present descriptive work. These include the formalization of architectural constraints for AI-assisted development, governance-as-code frameworks, certification and regulatory alignment, and economic models of governance investment. We have addressed these only briefly in the paper, but each warrants its own treatment in the form of focused subsequent research.

Trustworthy autonomy in software-intensive systems incorporating generative AI is not a destination on a single road. It is a configuration to be aligned, in a context that determines what alignment means. The empirical contribution of this paper is to make that configuration visible and assessable; the conceptual contribution is to argue that the multidimensional configurational view is the right level of analytical resolution. The fidelity is what the field now needs, and what the practitioners we interviewed are operating within whether their formal frameworks recognize it or not.

\begin{acknowledgements}
We would like to thank the anonymous interviewees for their valuable contributions to their research. Also, we thank Software Center \footnote{www.software-center.se} for supporting this research. Finally, the genesis of this paper was created during Shonan meeting \#241.

The authors used Claude Opus 4.7 to support qualitative coding and per-interview summarization during data analysis, as described in Section~\ref{resmet}. AI tools were not used to generate scientific content, claims, or interpretations; all conceptual contributions, the dimensional model, and the analytical findings reported in this paper are the work of the authors. The authors take full responsibility for the content of the publication.
\end{acknowledgements}

%% =====================================================================
%% Declarations
%% Place immediately before the bibliography (\bibliography{...} or
%% \begin{thebibliography}{...}). The acknowledgements environment can
%% remain separately, or its content can be folded into Funding below.
%% =====================================================================

\section*{Declarations}

\paragraph*{Funding.}
This research was supported by Software Center
, a collaboration between Swedish industry
partners and academic institutions including Chalmers University of
Technology and Malmö University (\url{www.software-center.se}). It was also supported by the U.S.\
National Science Foundation, under grant number~2232721. The funders had
no role in study design, data collection, analysis, interpretation, or
the decision to submit this work for publication.

\paragraph*{Ethical approval.}
This study consisted of semi-structured interviews with consenting adult
professionals about their organizations' practices and experiences with
generative AI in software engineering. The research did not involve
patients, vulnerable populations, sensitive personal data beyond the
professional context, or any clinical or behavioral intervention. In
accordance with the applicable research-ethics frameworks at the
authors' institutions, expert interview studies of this kind do not
require formal ethics-committee approval. All interviews were conducted
under standard research-ethics principles, including voluntary
participation, informed consent, the right to withdraw at any time, and
confidentiality of the participants and their organizations.

\paragraph*{Informed consent.}
Informed consent was obtained from all individual participants included
in the study. Prior to each interview, participants were informed of the
purpose of the research, the nature of the questions, the expected
duration, the intended use of the recorded material, the anonymization
protocol, and their right to withdraw at any time without consequence.
Participants consented to the audio recording and automated
transcription of the interview. No personally identifying information
about participants or their organizations appears in the published
paper; respondents are referred to by anonymized codes (P1--P18), and
organizations are characterized by domain descriptors sufficient to
preserve analytical context.

\paragraph*{Author contributions.}
All authors contributed to the conception, design, and execution of the
study. Interview protocol design was a joint effort by all four authors.
Data collection was distributed among all authors: Jan Bosch and Helena
Holmström Olsson conducted the interviews with respondents P1--P4, P6,
P9, P18; Henry Muccini conducted the interviews with respondents P5 and
P17; Rick Kazman conducted the interviews with respondents P7, P8,
P10--P16. Each interviewing team performed initial coding of its own
interviews. Cross-corpus axial coding, synthesis of the trust paradigms,
and formalization of the eight-dimensional health check model were
carried out collaboratively by all four authors through iterative shared
analysis. Jan Bosch led the drafting of the manuscript, with substantive
contributions to writing, revision, and intellectual content from Rick
Kazman, Henry Muccini, and Helena Holmström Olsson. All authors read and
approved the final manuscript, and all authors take responsibility for
its content.

\paragraph*{Data availability.}
The interview transcripts and per-interview codings that support the
findings of this study are not publicly available due to confidentiality
commitments made to the participating practitioners and their
organizations. Several respondents disclosed information about internal
practices, ongoing projects, and organizational decisions that they
consented to share under the condition that the raw material remain
restricted to the research team. The anonymized analytical synthesis,
including all verbatim quotations reported in the paper, is available
within the published article. Researchers seeking to build on this work
are encouraged to contact the corresponding author to discuss possible
collaboration or methodological replication; the interview protocol used
in this study can be made available on reasonable request.

\paragraph*{Conflict of interest.}
The authors maintain active industry consulting and research
partnerships in domains overlapping with those of several interviewees,
as disclosed in Section~\ref{sec:positionality} of the manuscript. To
mitigate any influence on findings, no respondent was interviewed about
a project on which the interviewing author had an ongoing consulting
relationship, and member-checking communication was restricted to
confirmation of analytical synthesis. The authors declare no other
competing financial or non-financial interests relevant to the content
of this article.

\paragraph*{Clinical trial number.}
Not applicable.

% BibTeX users please use one of
%\bibliographystyle{spbasic}      % basic style, author-year citations
%\bibliographystyle{spmpsci}      % mathematics and physical sciences
%\bibliographystyle{spphys}       % APS-like style for physics
%\bibliography{}   % name your BibTeX data base

% Non-BibTeX users please use
\bibliographystyle{spbasic}
\bibliography{bibliography}

\end{document}